\documentclass[a4paper,11pt]{article}
\usepackage{epsfig}
\usepackage{amsfonts}
\usepackage{latexsym}
\usepackage{color}
\usepackage{amsfonts}
\usepackage{graphicx}
\usepackage[colorlinks]{hyperref}
\addtocounter{section}{0}
\renewcommand{\theequation}{\thesection.\arabic{equation}}

\newcommand{\ba}{\begin{array}}
\newcommand{\ea}{\end{array}}
\newcommand{\be}{\begin{equation}}
\newcommand{\ee}{\end{equation}}
\newcommand{\nn}{\nonumber}
\newcommand{\bea}{\begin{eqnarray}}
\newcommand{\eea}{\end{eqnarray}}
\newcommand{\beas}{\begin{eqnarray*}}
\newcommand{\eeas}{\end{eqnarray*}}

\begin{document}

\begin{center}
\begin{Large} {\Large{
New solutions to the $sl_q(2)$-invariant Yang-Baxter equations at
roots of unity}}
\end{Large}\vspace{0.2cm}\\ {D. Karakhanyan}, {Sh. Khachatryan} \\
\begin{small}{Yerevan Physics Institute after A. Alikhanyan,\\ Br. Alikhanian 2, Yerevan
36, Armenia}\end{small}
\end{center}

 We find
 new solutions to the Yang-Baxter equations with the
 $R$-matrices possessing $sl_q(2)$ symmetry
 at roots of unity,
 using indecomposable representations.
 The corresponding
 quantum one-dimensional chain models, which can
 be treated as extensions of the XXZ model at roots of unity, are
 investigated.
 We consider the case $q^4=1$. The
 Hamiltonian operators of these models as a rule appear to be 
 non-Hermitian. Taking
 into account the correspondence between the representations
 of the quantum algebra $sl_q(2)$ and the quantum super-algebra
 $osp_t(1|2)$, the presented analysis
 can be extended to the latter
 case for the appropriate values of the deformation parameter.
\newpage
\section{Introduction}

The solutions to the Yang-Baxter equations (YBE) for the quantum
algebra $sl_q(2)$ \cite{ybe,ks} when deformation parameter $q$ is
given by a root of unity \cite{s1,ak,ps} are widely investigated
for irreducible ("spin", (semi-)cyclic and nilpotent)
representations \cite{BS, RGS}. In this work we would like to fill
up the existing gap by considering indecomposable ones
\cite{ak,ps,kkh}. We
show that use 
of these representations 
 provides a large
number of new solutions to the YBE and correspondingly a rich
variety of the $sl_q(2)$-invariant integrable models 
at roots of unity.

The solutions to the YBE with the given symmetry admit linear
decomposition over the symmetry-invariant objects - projectors
\cite{krs,j}. Our strategy in looking for a new solution to the
Yang-Baxter equations is straightforward. After substitution of
the most general linear combination of the appropriate
$sl_q(2)$-invariant objects
(projectors) 
into the YB equations, 
 the latter ones are reduced to the set of the functional equations
defined on the corresponding coefficients. At roots of unity it
takes place a degeneration of the standard fusion rules of the
quantum algebras, and it introduces some modifications in the
formulation of the ${R}_{A'A''}$-matrices, defined on the tensor
product of two
spaces, $A'\otimes A''$, in terms of the projectors. In this paper we 
consider 
 the highest and lowest weight representations of the
quantum algebra when $q$ is a root of unity, and the analysis is
restricted to the representations, which have their analogues at
general $q$ or are emerging from their fusions (so-called $A$-type
representations \cite{s1,ak,ps}). They are grouped into two
classes: irreducible spin-representations $V$ (spin-irrep) and
corresponding indecomposable representations $\mathcal{I}$. So the
task is to define the structure of the $R_{VV}$-,
$R_{V\mathcal{I}}$- and $R_{\mathcal{I}\mathcal{I}}$-matrices in
terms of the projection operators, obtaining preliminarily all the
variety of the projectors.  At roots of unity the number of the
projectors acting on the spaces of the tensor products
$\mathcal{I}\otimes V$ or $\mathcal{I}' \otimes \mathcal{I}''$
becomes larger than the number of the projectors in the case of
general $q$ (when instead of $\mathcal{I}$ a direct sum of two
irreps stands), and it leads to the increasing of the number of
the solutions to the YBE. The obtained solutions 
allow us to construct new integrable models with Hamiltonian
operators invariant with respect to the mentioned quantum algebra
at roots of unity. New solutions are found in this paper,
particularly, for the case $q^2=-1$. By means of them  quantum 
integrable chain models are constructed with the fundamental
spin-1/2 representations  on the sites, using the fact, that
four-dimensional indecomposable representation is a direct product
of two spin-1/2 irreps.

Investigation of the solutions to the YBE using the $B$-type
representations (including cyclic, semi-cyclic and nilpotent
irreps and corresponding indecomposable representations), which
have no their analogues at general $q$ \cite{s1,ak,ps}, will be
done afterwards.

A similar analysis would be valid also for the case of the
quantum super-algebra $osp_t(1|2)$ \cite{k1,kr,s,k,sca}, 
due to the existing correspondence between the representations
 of the quantum algebras $sl_q(2)$ and
 $osp_t(1|2)$ with $q=it^{1/2}$ \cite{z,k,kkh,SHKKH}. Note, however, that
 when $q=\pm i$ ($t=1$) the mentioned correspondence  does not take place,
 because the non-deformed super-algebra $osp(1|2)$ has no even-dimensional representations. 

The paper is organized as follows: in the first section we review
the known ways to find solutions to the YBE. The second and third
sections are devoted correspondingly to the description of the new
solutions found for the exceptional values of the deformation
parameter $q$ and to the construction of the corresponding
integrable chain models. The YB equations at this
case have a huge number of the solutions. We 
discuss  three large classes of the solutions in Section 2.
In Section 3 we 
consider some of the Hamiltonian operators corresponding to
the obtained $R$-matrices chosen as (symmetric) representatives 
of  each class of the solutions, displaying the variety of the
resulting 1d quantum chain models.
 The fourth section briefly depicts the
character of the dynamics of the systems possessing non-Hermitian
and non-diagonalizable Hamiltonian operators which met in the
third section. In the Appendix the projection operators are
described in general terms (an addition to Sections 1.2 and 1.3),
and for $q=i$, particularly.

%
%

\subsection{ $sl_q(2)$ algebra  and Jimbo's relations for composite
$R$-matrices.}

We define the algebra relations and co-product for quantum algebra
$sl_q(2)$ as
\bea &[e,f]=\frac{k-k^{-1}}{q-q^{-1}},\quad  q^2 e k= k e,\quad f
k=q^2 k f,&\\& \Delta[e]=e\otimes k^{-1/2}+ k^{1/2}\otimes e,\;\;
\Delta[f]=f\otimes k^{-1/2}+k^{1/2} \otimes f,\;\;
\Delta[k]=k\otimes k,&\\& R \Delta=\bar{\Delta} R.&\label{algebra}
\eea
 Here $R$ is an intertwiner matrix characteristic to the quasi-triangular
Hopf algebra, and $\bar{\Delta}=P \Delta P$, where
 $P$ is a permutation operator $P:A'\otimes A''=A''\otimes A'$. The co-product $\Delta$ is a
 co-associative operation:
 $\Delta(1\otimes \Delta)=\Delta(\Delta\otimes 1)$.
The intertwiner matrix $R$ satisfies to the constant Yang-Baxter
equation
\be R_{12}R_{13}R_{23}=R_{23}R_{13}R_{12}. \ee
 $R_{ij}$ acts on the tensor product of two
representation spaces of the algebra, $A_i\otimes A_j$.
Irreducible representations of  $sl_q(2)$ at general $q$ are
classified similar to the spin-irreps of the non-deformed algebra
$sl(2)$: $r$-dimensional irrep $V_r$ is characterized by the spin
value $j=(r-1)/2$. The quadratic Casimir operator, defined as
\be c=f e+(q k+q^{-1} k^{-1})/(q-q^{-1})^2, \label{cas}\ee
 has the eigenvalue
$[r/2]_q^2+\frac{2}{(q-q^{-1})^2}$ on $V_r$. The tensor product of
two irreps has linear decomposition,
\bea V_{r_1}\otimes V_{r_2}=\bigoplus_{r=|r_2-r_1|+1}^{r_2+r_1-1}
V_r,\qquad \triangle r=2. \label{ccf}\eea
In this paper we 
denote the Casimir operator $c$
acting on the space $V_{r_1}\otimes V_{r_2}\otimes \cdots V_{r_p}$
also as $c^{r_1 r_2\cdots r_p}$.

 In the theory of the integrable models  the solutions $R_{ij}(u)$
 to the Yang-Baxter equations
with spectral parameter \cite{rbaxter}, 
\be
R_{12}(u-v)R_{13}(u)R_{23}(v)=R_{23}(v)R_{13}(u)R_{12}(u-v),\label{rrr}
\ee
 acquire an important role. The solutions of (\ref{rrr}) are
defined up to the following multiplicative transformations:
$R_{ij}(u)\to f(u) R_{ij}(a u)$, with arbitrary number $a$ and
arbitrary function $f(u)$. Jimbo's construction gives an
opportunity to derive solutions to (\ref{rrr}) from algebraic
relations \cite{j,kr,s}. In the work \cite{j} the author 
stated that Eq. (\ref{rrr}) must be satisfied, if the matrix
$R_{ij}(u)$ obeys
 the relations
 \bea\nn&\check{R}(u)\left( q^u f \otimes k^{1/2} +q^{-u} k^{-1/2} \otimes f
\right)=\quad\quad\quad\quad\quad\quad\quad\quad\quad\quad&\\\nn&
\quad\quad\quad\quad\quad\quad\quad\quad\quad\quad=\left(q^{-u}f\otimes
k^{1/2}+q^u k^{-1/2}\otimes f\right)\check{R}(u),&\\\label{jimbo}
&\check{R}(u)\left(q^u k^{-1/2}\otimes e+q^{-u} e\otimes k^{1/2}
\right)=\quad\quad\quad\quad\quad\quad\quad\quad\quad\quad&\\\nn
&\quad\quad\quad\quad\quad\quad\quad\quad\quad\quad =\left(q^{-u}
k^{-1/2}\otimes e+q^u e\otimes k^{1/2}\right)\check{R}(u).& \eea
Here $\check{R}(u)=PR(u)$,  for which
\be [\check{R}(u),\Delta]=0. \label{commuta}\ee

 When $q^n=1$ \cite{s1,ak,ps,grs}, then the number of the permissible irreducible
representations is restricted: the irreps $V_r$ can be of
dimensions $r=1,...,\mathcal{N}$, where $\mathcal{N}=n$, if $n$ is
odd and $\mathcal{N}=n/2$, if $n$ is even. The center of the
algebra is enlarged, new Casimir operators appear, which are
$e^{\mathcal{N}},\; f^{\mathcal{N}}$ and $k^{\mathcal{N}}$. The
irreducible representations are grouped into two types: $A$-type,
which includes ordinary spin representations ($e^{\mathcal{N}}=
0,\; f^{\mathcal{N}}=0$ and $k^{\mathcal{N}}= \pm 1$) with
dimensions $\leq \mathcal{N}$ , and $B$-type, which consists of
cyclic ($e^{\mathcal{N}}\neq 0,\; f^{\mathcal{N}}\neq 0$),
semi-cyclic ($e^{\mathcal{N}}\neq 0,\; f^{\mathcal{N}}=0$ or
$e^{\mathcal{N}}= 0,\; f^{\mathcal{N}}\neq 0$) and nilpotent
representations ($e^{\mathcal{N}}= 0,\; f^{\mathcal{N}}=0$ and
$k^{\mathcal{N}}\neq \pm 1$) with dimensions equal to
$\mathcal{N}$.

 Among the non-reducible representations
of the quantum algebra together with the irreducible 
representations 
there are also indecomposable ones, 
$\mathcal{I}_{A/B}$, of dimension
$\mathcal{R}=2 \mathcal{N}$ 
\cite{s1,ak,ps,kkh,s,sca,SHKKH}.
 It is known that $A$-type representations are forming a closed
fusion ring \cite{s1,ak,kkh}. We borrow from the work \cite{kkh}
the notations for $A$-type indecomposable representations,
$\mathcal{I}^{(\mathcal{R})}_{\{r,\mathcal{R}-r\}}$, where 
 $r$ ($r>\mathcal{N}$) is the dimension of the maximal proper subspace of
$\mathcal{I}^{(\mathcal{R})}_{\{r,\mathcal{R}-r\}}$, denoted
below by an abstract notation $\mathcal{U}$: it has
$(\mathcal{R}-r)$-dimensional proper irreducible subspace $U$. In
the fusions indecomposable representation
$\mathcal{I}^{(\mathcal{R})}_{\{r,\mathcal{R}-r\}}$ arises from
the "merging" of the representations $V_{r}$ and
$V_{\mathcal{R}-r}$ at roots of unity, when
$c_r=c_{\mathcal{R}-r}$ and $V_{r}\Rightarrow\mathcal{U},\;
V_{\mathcal{R}-r}\Rightarrow U$ (see for details
\cite{s1,ak,kkh}).

 We have excluded from the
present consideration the
 highest/lowest weight nilpotent
representations, for which $k^{\mathcal{N}}$ is generic. But of
course, such kind of investigation, which is done in this work,
can be carried out for them as well, previously making  proper
changes in the definitions 
of the indecomposable
representations, as the representations in this case are
parameterized by a continuous parameter (the value of
$k^{\mathcal{N}}$). Also
all the  representations of $B$-type 
 can
be considered. As it is known the fusion of the $B$-type
representations
can contain the indecomposable representations of $A$-type  \cite{ak}. 
Therefore the investigation of the solutions to the  YBE 
for the representations of $B$-type will include the results of
this paper particularly.  All these questions
we are addressing to  our 
subsequent 
 investigations.

In order to write down equations for indecomposable
representations, similar to Eqs. (\ref{jimbo}), which lead to a
simpler set of algebraic equations instead of the functional ones,
let us write the Yang-Baxter equations with  Lax operator $L$
\cite{j} (below $r_i$ denotes the dimension of the representation,
on which the operator acts):
\bea R^{r_1 r_2}(u-v) L^{r_1}(u) L^{r_2}(v) =L^{r_2}(v)
L^{r_1}(u)R^{r_1 r_2}(u-v),\label{ybe1}\eea
where $L^{r}$ is $2\times 2$ matrix
 with operator-valued
  elements acting on the space $V_r$
 \bea
&L^{r}(u)=q^u L_{+}-q^{-u}L_{-}\;,\qquad L_{+}=\left(\ba{cc}
k^{1/2}&g_{f}f
\\0&k^{-1/2}\ea\right),\label{lpm}
\quad L_{-}=\left(\ba{cc} k^{-1/2}&0
\\g_{e}e&k^{1/2}\ea\right).&
 \eea
We take $g_{f}=\frac{q^2-1}{q^{3/2}}$ and
$g_e=\frac{1-q^2}{q^{1/2}}$. The relations (\ref{jimbo}) can be
obtained from the equation (\ref{ybe1}), expanding r.h.s. and
l.h.s. of the latter in 
powers of $q^v$ and taking the expressions linear in respect of
$q^v$ (or $q^{-v}$). In the case, when one of the representations,
on which $R_{12}$ acts, say the second one, is a composite one
(i.e. can be represented as $V_{r_{2}'}\otimes V_{r_2''}$), then
$L^{r_2}$ must be modified. A natural generalization is to replace
the algebra generators $e,\;f,\; k$ in the expression (\ref{lpm})
of $L^{r_2}$ by the co-products
$\Delta[e],\;\Delta[f],\; \Delta[k]$. It will give 
$\check{R}^{r_1\;r_2'\times r_2''}$-matrix, which after
multiplication from the left and right sides by proper projectors
$1\otimes P^r$ ($P^r\otimes 1$), becomes $R^{r_1\;{r}}$, where
$(|r_1-r_2|+1)\leq r\leq (r_1+r_2-1)$. We do not consider the
possibility of $(P^{r'}\otimes 1)\check{R}^{r_1\;r_2'\times
r_2''}(1\otimes P^{r''})$, with $r'\neq r''$, as the
$\check{R}$-matrices are defined so that they are commuting with
the algebra generators (\ref{commuta}).

If we want to take into account the entire space of the fusion
representations, we can write down $L^{ r_2'\times r_2''}$ as the
following tensor product $L^{ r_2'}(u)\otimes L^{r_2''}(w)$.
\bea\nn & \check{R}^{r_1\; r_2'\times
r_2''}(u\!-\!v,u\!-\!w)L^{r_1}(u)\left[L^{r_2'}(v)\otimes
L^{r_2''}(w)\right]=\left[L^{r_2'}(v)\otimes
L^{r_2''}(w)\right]L^{r_1}(u) \check{R}^{r_1\;r'_2\times
r_2''}(u\!-\!v,u\!-\!w).& \nn\\\label{ybecc}\eea

Besides of the usual commutativity relations $\check{R}^{r_1\;
r_2'\times r_2''}
\Delta(\Delta[a])=\Delta(\Delta[a])\check{R}^{r_1\; r_2'\times
r_2''}$, $a=e,f,k^{\pm}$, the non-diagonal elements of the
matrix-relations (\ref{ybecc}) contain also spectral parameter
dependent relations, which are more complicated than
(\ref{jimbo}): we shall refer to them as Jimbo's relations for
composite (including tensor products of the irreps)
representations. Here we write the following equations for the
generator $f$ (we suppose $v=w$ in (\ref{ybecc}), and
$\check{R}^{r_1\; r_2'\times r_2''}(u,u)\equiv \check{R}(u)$)
\bea \label{jc}\!\!\!\!\!\!\!\!\!\!\!\!\!\!\!
&\!\!\!\!\!\check{R}(u)\Big(\!q^u (\Delta[f]\otimes
k^{\frac{1}{2}}\! +\!k^{\frac{1}{2}}\otimes
k^{\!-\frac{1}{2}}\otimes f\!-\!\frac{(1-q^2)^2}{q^{2}}f\otimes e
\otimes f\! +\!f \otimes k^{\frac{1}{2}}\otimes
k^{\!-\frac{1}{2}})\!+\!q^{\!-u}k^{\!-\frac{1}{2}}\otimes\Delta[f]\!\Big)\!\!\!\!\!\!&\\&\!\!\!\!\!
=\!\Big(q^{-u}\Delta[f]\otimes k^{\frac{1}{2}}\! +\!q^u(
k^{-\frac{1}{2}}\otimes \Delta[f]\!+\!k^{\frac{1}{2}}\otimes
k^{\!-\frac{1}{2}} \otimes f\!-\!\frac{(1-q^2)^2}{q^{2}}f\otimes e
\otimes f\!+\!f\otimes k^{\frac{1}{2}}\otimes
k^{\!-\frac{1}{2}})\Big)\check{R}(u).&\nn\eea
and
\bea \label{jimbo_comp}&\check{R}(u)\Big(q^u f\otimes k^{
\frac{1}{2}}\otimes k^{\frac{1}{2}} +q^{-u}(k^{-\frac{1}{2}}
\otimes f\otimes k^{\frac{1}{2}} +k^{-\frac{1}{2}}\otimes
k^{-\frac{1}{2}}\otimes f)\Big)&\\&=\Big(q^u
k^{-\frac{1}{2}}\otimes k^{-\frac{1}{2}}\otimes f +q^{-u}(
k^{-\frac{1}{2}}\otimes f\otimes k^{\frac{1}{2}}+f\otimes
k^{\frac{1}{2}} \otimes k^{\frac{1}{2}})\Big)\check{R}(u).&
\nn\eea
In case of $v\neq w$ in (\ref{ybecc}), the equations derived above
 contain the parameter $v-w=u_0$; e.g. the last relation
 takes the form
\bea \label{jimbo_comp1}&\check{R}(u,u+u_0)\Big(\!q^{u+u_0}
f\otimes k^{ \frac{1}{2}}\otimes k^{\frac{1}{2}}
+q^{-u}(q^{u_0}k^{-\frac{1}{2}} \otimes f\otimes k^{\frac{1}{2}}
+q^{-u_0}k^{-\frac{1}{2}}\otimes k^{-\frac{1}{2}}\otimes
f)\!\Big)&\\&=\Big(\!q^{u+u_0} k^{-\frac{1}{2}}\otimes
k^{-\frac{1}{2}}\otimes f +q^{-u}( q^{-u_0}k^{-\frac{1}{2}}\otimes
f\otimes k^{\frac{1}{2}}+q^{u_0}f\otimes k^{\frac{1}{2}} \otimes
k^{\frac{1}{2}})\!\Big)\check{R}(u,u+u_0).& \nn \eea

The extension of these equations for the matrices $R^{r'_1\times
r''_1 \; r'_2\times r''_2}$ acting on the space $[V_{r'_1}\otimes
V_{r''_1}]\otimes[V_{r'_2}\otimes V_{r''_2}]$ can be found taking
$L^{r'_1}\otimes L^{r''_1}$ instead of $L^{r_1}$ in (\ref{ybecc}).

\subsection{Projection operators and indecomposable representations.}

At general values of $q$ the tensor product $V_{r_1}\otimes
V_{r_2}$ admits Clebsh-Gordan decomposition (\ref{ccf}),
 and the eigenvalues $c_r$ of
the Casimir operator $c$ are different for different $r$. It
means, that any invariant operator $a,\; [a,g]=0, \;g\in sl_q(2)$,
acts on each of the irreducible spaces as an identity operator,
and hence can be represented as a sum over the projection
operators $P_r$ on these spaces:
\be a=\sum_r a_r P_r,\;\;\;  P_r P_{r'}=P_r \delta_{r r'}.
\label{cg}\ee
Particularly, $c=\sum_{r=|r_1-r_2|}^{r_1+r_2-1}c_r P_r$. This
means, that $\check{R}^{r_1 r_2}$-matrix 
 ($\check{R}^{r_1 r_2} : V_{r_1}\otimes
V_{r_2}\Rightarrow V_{r_2}\otimes V_{r_1}$; when $r_1\neq r_2$,
the relation (\ref{commuta}) implies $\check{R}^{r_1
r_2}\Delta^{r_1\;r_2}=\Delta^{r_2\;r_1}\check{R}^{r_1 r_2}$)
acquires the form $\check{R}^{r_1 r_2}(u)=\sum_{
r=|r_1-r_2|}^{r_1+r_2-1}f_r(u) \breve{P}_r$ \cite{j,krs,SHKKH}.
Here $\breve{P}_r\equiv \it{\mathcal{P}}^{r_1 r_2} P_r$, with
$\it{\mathcal{P}}^{r_1 r_2}$ being an identical transformation
operator translating the space $V_{r_1}\otimes V_{r_2}$ into the
isomorphic space $V_{r_2}\otimes V_{r_1}$, and
$\it{\mathcal{P}}^{r\; r'}\it{\mathcal{P}}^{r'\;
r}=\mathbb{I}^{r\;r}$, $\it{\mathcal{P}}^{r\; r}=\mathbb{I}$
($\mathbb{I}$ is the unity operator defined on the space
$V_{r}\otimes V_{r}$).

When
 at least one of the representations $V_{r_1}$ and $V_{r_2}$
is not irreducible, then in the decomposition of their tensor
product some irreps have the same eigenvalues of the Casimir
operator. Suppose, $R^{r\; r'}(u)$ acts on the tensor product
$U_r\otimes U_{r'}$, where $U_r$ or/and $U_{r'}$ are reducible,
and it takes place the fusion $U_r\otimes
U_{r'}=\bigoplus_{\bar{r}}
\bigoplus_i^{\epsilon_{\bar{r}}}V_{\bar{r}}^i$.
$\epsilon_{\bar{r}}$ is the multiplicity of the irrep
$V_{\bar{r}}$, $\sum_{\bar{r}} \epsilon_{\bar{r}}=r r'$. Here 
an additional index $i \in\{1,...,\epsilon_{\bar{r}}\}$ is
attached to distinguish isomorphic irreps $V_{\bar{r}}^i$
corresponding to the same eigenvalue $c_{\bar {r}}$. Then among
the invariant operators, commuting with the algebra generators,
also projectors $P_{\bar{r}}^{ij}$ appear, which map irreps
$V_{\bar {r}}^i$ to each other. So, the $R$-matrix, as any
invariant operator, admits a linear representation over the set of
the projectors $P_{\bar{r}}^{ij}$ of number
$\sum_{\bar{r}}\epsilon_{\bar{r}}^2$, i.e.
\bea \check{R}^{r\; r'}(u)=\it{\mathcal{P}}^{r\;
r'}\sum_{\bar{r}}\sum_{i,j} f_{\bar{r}}^{ij}(u)
P_{\bar{r}}^{ij},\qquad\qquad\;P_{\bar{r}}^{ij} P_{\bar{r}'}^{k
r}=P_{\bar{r}}^{i r}\delta_{j k}\delta_{\bar{r} \bar{r}'}.\eea

At the exceptional values of deformation parameter $q$, as it was
stated, among the representations on which the $R$-matrix acts
also indecomposable representations $\mathcal{I}$ can be included
along with the ordinary irreducible representations $V$. In this
case the set of the possible projectors includes also the
operators $P':\mathcal{I}\to \mathcal{I}$, which are acting inside
of the spaces of the indecomposable representations not as  unity
matrices. The symbolic structure of the indecomposable
representation can be shown as $\mathcal{I}=\mathcal{U}\cup
\mathcal{U}'$, on which the algebra generators $\{g\}$ act in the
following way
\bea g\cdot \mathcal{U}\Rightarrow\mathcal{U},\;\;\; g\cdot
\mathcal{U}'\Rightarrow\mathcal{I}. \eea
  The vectors belonging to
$\mathcal{U}'$ are defined up to the addition of the vectors
belonging to an irreducible representation $U$
($\mbox{dim}[\mathcal{U}']=\mbox{dim}[U]$), which is the proper
subspace of $\mathcal{U}$ and have vectors with zero norm
\cite{s,kkh}. The action of the Casimir operator on these spaces
is given by: $c\cdot\mathcal{U}= c_{\mathcal{I}}\mathbb {I}\cdot
\mathcal{U}$, where $\mathbf{I}$ is the unit operator, and $c
\cdot \mathcal{U}'= c_{\mathcal{I}} \mathbb{I}\cdot \mathcal{U}' +
c_{\mathcal{I}}' \mathbb{I}\cdot U$. Similarly, together with the
usual $P$, acting as unity operator on the indecomposable
representation, a projection operator $P'$, $P'\cdot
\mathcal{U}=0,\; P'\cdot \mathcal{U}'= U$, can be introduced. In
the case, when decomposition includes $n\geq2$ isomorphic
indecomposable representations $\mathcal{I}^i=\mathcal{U}^i \cup
{\mathcal{U}'}^i$, one is able to construct $2n^2$ independent 
projection operators $P^{ij},\; {P'}^{ij}$, $i,j=1,...,n$, acting
as
\bea\label{ppi} &P^{ij}\cdot \mathcal{I}^k = \delta_{jk}\mathcal{I}^i,&\\
&{P'}^{ij}\cdot \mathcal{U}^{k'} = \delta_{jk}\mathcal{
U}^i,\;\;\; {P'}^{ij}\cdot \mathcal{U}^k = 0.&\nn\eea
The projectors have the following obvious properties
\bea P^{ij} P^{kp}\!=\! P^{ip} \delta_{jk},\!\quad
{P'}^{ij}{P'}^{kp}\!\!\!=0,\;\;\;
P^{ij}{P'}^{kp}\!\!\!={P'}^{ij}P^{kp}\!.\label{projec}\eea
Note, that the isomorphic representations having the same
dimension, structure and eigenvalues of the Casimir operator, can
differ by the signs of the eigenvalues of the generator $k$,
conditioned by the algebra automorphism $k \to -k,\;\; e \to \pm
e,\;\;\; f\to \mp  f$.
 The projectors $P^{ij}$ and ${P'}^{ij}$ relate to each other only
vectors with the same set of the eigenvalues of $k$, as it is
implied by symmetry. And it means, that for the mentioned
situation the action of the projectors $P^{ij},\; {P'}^{ij}$ must
have slight modification in comparison with  (\ref{ppi}). We shall
touch all these aspects in details below for the discussed cases.

\subsection{Projectors and Casimir operator.}
In this subsection we want to present another approach to the
problem. Let we are given a set of the algebra representations
$\mathcal{S}=\{V,\mathcal{I}\}$ and let us consider on this set a
general matrix, which is commutative with the algebra. The number
of degrees of freedom of this matrix is given by the number of the
mutually linear independent matrices (basis matrices) which are
invariant with respect to the symmetry algebra. We can choose as
the basis matrices the projection operators described above, i.e.
the operators which act non-trivially (are not zero) only on one
non-reducible space, mapping the latter either to itself or to
another non-reducible space. Note, that each invariant operator
on $\mathcal{S}$, including the identity and Casimir operators,
can be represented as a linear superposition of these operators.
Now we discuss the inverse problem: how the projection operators
can be built by means of the Casimir and unity operators.

The case (\ref{cg}) discussed in the beginning of the previous
section corresponds to $\mathcal{S}=V_{r_1}\otimes V_{r_2}$ (
\ref{ccf}), and the projectors $P_r$, as it is well known, are
given by polynomials of degree $r_1 +r_2 -1$ in terms of the
Casimir operator $c$, as the eigenvalues $c_r$ at general $q$ do
not coincide one with other:
\bea P_r=\prod_{p\neq r}\frac{c-c_p
\mathbb{I}}{c_r-c_p}.\label{pc}\eea
Let us now consider some particular cases, when $\mathcal{S}$
contains indecomposable representations. If it consists of a
single indecomposable representation $\mathcal{S}=\mathcal {I}$,
then
 \bea
c=c_{\mathcal{I}} P_{\mathcal{I}}+c'_{\mathcal{I}} P'_{\mathcal{I}}
,\qquad\qquad\; P_{\mathcal{I}}=\mathbb{I},\qquad P'_{\mathcal{I}}=
\frac {c-c_{\mathcal{I}}\mathbb{I}}{c'_{\mathcal{I}}}.
 \eea
When $\mathcal{S}=\mathcal{I}\oplus V_r$, one has
\bea
&c=c_{\mathcal{I}}P_{\mathcal{I}}+c'_{\mathcal{I}}P'_{\mathcal
{I}}+c_r P_r,\qquad \mathbb{I}= P_{\mathcal{I}}+ P_r,&\\\nn
&P'_{\mathcal{I}}=\left(\frac{c-c_{\mathcal{I}}\mathbb{I}}{c'_{
\mathcal{I}}}\right)\left(\frac{c-c_{r}\mathbb{I}}{c_{\mathcal{I}}-
c_r}\right),&\\\nn &P_{\mathcal{I}}=\left(\frac{c-(2 c_{\mathcal
{I}}-c_r)\mathbb{I}}{c_r-c_{\mathcal{I}}}\right) \left(\frac{c-
c_{r}\mathbb{I}}{c_{\mathcal{I}}-c_r}\right),\qquad P_r=\left(\frac
{c-c_{\mathcal{I}}\mathbb{I}}{c_r-c_{\mathcal{I}}}\right)^2.&
 \eea
The next simple case is $\mathcal{S}=\mathcal{I}_1\oplus
\mathcal{I}_2$, $c_{\mathcal{I}_1} \neq c_{\mathcal{I}_2}$. Here
the following formulas take place:
\bea &c=c_{\mathcal{I}_1} P_{\mathcal{I}_1}+c'_{\mathcal{I}_1}
P'_{\mathcal{I}_1}+c_{\mathcal{I}_2}
P_{\mathcal{I}_2}+c'_{\mathcal{I}_2} P'_{\mathcal{I}_2},&\\\nn
&P'_{\mathcal{I}_i}=\left(\frac{c-c_{\mathcal{I}_i}\mathbb{I}}{c'_{\mathcal{I}_i}}\right)
\left(\frac{c-c_{\mathcal{I}_j}\mathbb{I}}{c_{\mathcal{I}_i}-c_{\mathcal{I}_j}}\right)^2,\;\;\;
i=1,2,\;\; j\neq i,&\\\nn &P_{\mathcal{I}_i}=\left(\frac{2 c-(3
c_{\mathcal{I}_i}-c_{\mathcal{I}_j})\mathbb{I}}{c_{\mathcal{I}_j}-c_{\mathcal{I}_i}}\right)
\left(\frac{c-c_{\mathcal{I}_j}\mathbb{I}}{c_{\mathcal{I}_i}-c_{\mathcal{I}_j}}\right)^2\!\!\!,\;
i=1,2,\; j\neq i.&\eea
Above formulas have obvious generalizations for the set
$\mathcal{S}=V_{r_1}\oplus \cdots \oplus V_{r_n}\oplus
\mathcal{I}_{1}\oplus\cdots \oplus\mathcal{I}_{p}$, where all the
representations have different eigenvalues of 
$c$:
\bea &c=\sum_{i=1}^n c_{r_i}
P_{r_i}+\sum_{j=1}^p(c'_{\mathcal{I}_j}
P'_{\mathcal{I}_j}+c_{\mathcal{I}_j} P_{\mathcal{I}_j})&\\\nn
&P_{r_k}=\prod_{i\neq k}^n\left(
\frac{c-c_{r_i}\mathbb{I}}{c_{r_k}-c_{r_i}}\right)\prod_{j}^p\left(
\frac{c-c_{\mathcal{I}_j}\mathbb{I}}{c_{r_k}-c_{\mathcal{I}_j}}\right)^2,&
\\\nn
&P'_{\mathcal{I}_k}=\frac{c-c_{\mathcal{I}_k}\mathbb{I}}{c'_{\mathcal{I}_k}}\prod_{i}^n
\left(\frac{c-c_{r_i}\mathbb{I}}{c_{\mathcal{I}_k}-c_{r_i}}\right)\prod_{j\neq
k}^p\left(
\frac{c-c_{\mathcal{I}_j}\mathbb{I}}{c_{r_k}-c_{\mathcal{I}_j}}\right)^2,&\\\nn
&P_{\mathcal{I}_k}=\left(\mathrm{c}_{V\mathcal{I}}\;c-
{\bar{\mathrm{c}}}_{V\mathcal{I}} \mathbb{I}\right)\prod_{i}^n
\left(\frac{c-c_{r_i}\mathbb{I}}{c_{\mathcal{I}_k}-c_{r_i}}\right)\prod_{j\neq
k}^p\left(
\frac{c-c_{\mathcal{I}_j}\mathbb{I}}{c_{r_k}-c_{\mathcal{I}_j}}\right)^2,&\\\nn
&\mathrm{c}_{V\mathcal{I}}=
\sum_i^n\frac{1}{c_{r_i}-c_{\mathcal{I}_k}} +\sum_{j\neq
k}^p\frac{2}{c_{\mathcal{I}_j}-c_{\mathcal{I}_k}},\;{\bar{\mathrm{c}}}_{V\mathcal{I}}=
\mathrm{c}_{V\mathcal{I}} c_{\mathcal{I}_k}-1&. \label{pvi}\eea
How should be generalized the above formulas in case of degeneracy
of the Casimir operator? The answer seems to be  simple: when the
eigenvalues spectrum of $c$ has degeneracy of degree $n$ then one
should consider an operator $c^\frac1n$ instead of $c$
($(c^{\frac{1}{n}})^n=c$), eigenvalues' spectrum of which is not
degenerated and one can use the formula (\ref{pvi}), replacing $c$
with $c^{\frac{1}{n}}$ and with its eigenvalues. A detailed
consideration is placed in the Appendix.

\section{Solutions to the YBE}

\addtocounter{section}{0}\setcounter{equation}{0}

The solutions $\check{R}^{r_1 r_2}$ to the YBE, when $V_{r_1}$ and
$V_{r_2}$ are irreps, for the quantum super-algebra $osp_q(1|2)$
at general $q$ are considered in \cite{SHKKH}. As there is a full
one-to-one correspondence between the representations of two
quantum algebras at general $q$ \cite{k,z,kkh}, we can take the
solutions given there and verify, that after the appropriate
change of the quantum deformation parameter, and after removing
the signs connected with the gradings, we shall arrive at the
solutions to the YBE for $sl_q(2)$.

Let us briefly represent all the solutions to the YBE at general
$q$ for inhomogeneous spectral parameter dependent $\check{R}^{r_1
r_2}(u)$-matrix. From Jimbo's relations (\ref{jimbo}) one finds
(below $r_1=2j_1+1,\;r_2=2j_2+1$)
\bea &{\check{R}}^{(r_1
r_2)}(u)=\sum_{j=|j_1-j_2|}^{j_1+j_2}\mathbf{r}_j(u)\breve{P}_{2j+1},\label{rgi}&\\&
 \mathbf{r}_{j'}(u)
\label{rij} =\prod_{j=j'}^{j_1+j_2-1}\left[\Upsilon_{j_1
j_2}^j\frac{q^{u}- q^{-u}q^{2(
j'+1)}}{q^{-u}-q^{u}q^{2(j'+1)}}\right]\mathbf{r}_{j_1+j_2}(u),
&\\\label{g}&\Upsilon_{j_1
j_2}^j=q^{i_2-i_1}\frac{\alpha_{j_2}^{j-i_1}}{\alpha_{j_1}^{j-i_2}
}\frac{C\left(^{j_1\;\;\; j_2\;\;\;j}_{i_1\;
j-i_1\;j}\right)C\left(^{j_2\;\;\;\;\;\;\; j_1\;\;\;\;\;\;j+1} _{
i_2\;\;\; j+1-i_2\; j+1}\right)}{C\left(^{j_1\;\;\;\;\;\;
j_2\;\;\;\;\;\;j+1} _{i_1\;\; j+1-i_1\;\;
j+1}\right)C\left(^{j_2\;\;\; j_1\;\;\;\;j} _{i_2\;
j-i_2\;j}\right)}.&\eea
where the projector operators $\breve{P}_r,\;\breve{P}_r \cdot
V_{g}=\delta_{r g}V_g$, are acting as map $V_{2 j_1+1}\otimes V_{2
j_2+1}\to V_{2j_2+1}\otimes V_{2j_1+1}$. When $r_1=r_2$, then
$\check{P}_r=P_r$ and $\Upsilon_{j_1 j_2}^j=1$ \cite{j,krs,s}.
 By the notations $C\left(^{ j_1\;\;\;j_2\;\;\;j} _{ i_1\;\; i-i_1\;\;\; i}\right)$
 we have denoted the Clebsh-Gordan
coefficients and the parameters $\alpha_{j}^{i}$ are the matrix
elements of the algebra generator $e$ on the vector space
$V_{2j+1}=\{[v_{i}]_j,\;i= -j,-j+1,...,j\}$: $e\cdot
[v_{i}]_j=\alpha_{j}^{i} [v_{i+1}]_j$, $k\cdot
[v_{i}]_j=q^{2i}[v_{i}]_j$. The expression (\ref{g}) is the same
for all permissible values of $i_1$ and $i_2$ from the range
$-j_1\leq i_1,\leq j_1$, $-j_2\leq i_2\leq j_2$ (see
\cite{kkh,SHKKH}).

By means of  Jimbo's  ordinary relations (\ref{jimbo}) or the
relations for composite matrices (\ref{jc}, \ref{jimbo_comp})
 we can find solutions to the YBE with
$\check{R}^{r_1 r'_2\times r''_2}$ ($\check{R}^{r'_1\times r''_1
r'_2\times r''_2}$).  These relations are inherited from the Lax
representations of the YBE (\ref{ybe1}, \ref{ybecc}) and their
solutions can be obtained by the descendant procedure from the
fundamental solution $R^{2\;2}(u)$ \cite{krs}. By this reason, as
we shall see, at roots of unity solving all  Jimbo's relations
leads to the solutions being the limit cases of those existing at
general $q$ (like the fundamental solution). So, at roots of unity
for obtaining essentially new solutions to the YBE one must
consider
directly the YBE. 
Note, although,  that (as we shall see later on, in Section 2.2)
using only one pair of Jimbo's composite relations (namely,
(\ref{jimbo_comp}), and its analogue for the generator $e$) will
bring at roots of unity to some definite generalizations of the
solutions existing at general $q$.

At general $q$ also there are solutions to the YBE which do not
admit Lax representation (i.e. do not obey the relations
(\ref{ybe1})). When $r_1=r_2=3$ besides of the solution $
\check{R}_{1}^{3\; 3}(u)$, which can be obtained from the general
solution (\ref{rij}), there is a separate solution
$\check{R}_2^{3\; 3}(u)$, 
which does not admit descendant solutions $R^{3 r_i},\; R^{r_j
r_i}$ for higher $r_i$ (see \cite{kr}, \cite{SHKKH}). Below there
is done a multiplicative transformation of the spectral parameter
of $\check{R}_{1}^{33}(u)$ in comparison with (\ref{rij}), $u\to
-u/2$:
\bea {\check{R}_{1}^{33}(u)}\!=\!P_5\!+\!\frac{q^{4+u}\!
-\!1}{q^4\! -\! q^u} P_3\!+\! \frac{(q^{2\!+\!u}\!
-\!1)(q^{4\!+\!u}\! -\!1)}{( q^2\!-\! q^u)(q^4\! -\! q^u)}
P_1,\;\;\;
\check{R}_2^{33}(u)\!=\! P_5\!+\!\frac{q^4 q^u\! -\! 1}{q^4\! -\!
q^u} P_3\!+ \!\frac{q^6 q^u\! +\! 1}{q^6\! +\! q^u} P_1.
\label{3solr}\eea
Also there is another solution, which does not obey (\ref{jimbo}),
and which does not distinguish the projectors $P_5$ and $P_3$,
namely
\bea \label{ra}&\check{R}_{\pm}^{3\; 3}(u) = P_5+P_3+
\frac{a_{\pm} + q^u}{1+a_{\pm} q^u} P_1,&\\\nn&
a_{\pm}=\frac{-1}{2 q^4}\left(1 + 2 q^2 + q^4 + 2 q^6 + q^8\pm
(1+q^2+q^4)\sqrt{1 + 2 q^2-q^4+2 q^6+q^8}\right).&\eea
Note, that $a_{+}a_{-}=1$ and hence $\check{R}_{+}^{3\;
3}(u)=\check{R}_{-}^{3\;3}(-u)$. This solution belongs to the
series of the $R^{rr}$ solutions which admit "baxterized"
\cite{rbaxter} form $R=q^u R^+ +q^{-u}R^{-}$,
$$\check{R}^{r\; r}(u) = \mathbb{I}+ (\frac{a +
q^u}{1+a q^u}-1) P_1, \quad
a=\frac{i+\sqrt{-1+4/[r]_q^2}}{-i+\sqrt{-1+4/[r]_q^2}}.$$

 Here $\mathbb{I}$ is the $r^2\times r^2$ unity matrix defined on the space $V^r\times
 V^r$. There is no generalization $\check{R}^{r_1\; r_2}(u)$ for such matrices in the case of
 $r_1\neq r_2$. At $r=2$ (\ref{ra}) coincides with the fundamental
 solution in (\ref{rgi}).

\subsection{YBE solutions $\check{R}_{VV}$: $\check{R}^{33}(u)$ 
and some notes and statements.}

\paragraph{ Solutions at $q^3=\pm 1$.}
As an illustrative example we consider here the case
$\mathcal{N}=3$, which will provide us with the characteristic
properties of the solutions $\check{R}_{VV}$ 
at roots of unity.

 At $q^3=\pm 1$ the existing
non-reducible representations of the algebra $sl_q(2)$ are the
irreps $V_2$, $V_3$ (for the super-algebra $osp_q(1|2)$ the
fundamental representation is the $V_3$) and the indecomposable
representations $\mathcal{I}^{(6)}_{\{4,2\}}$ and
$\mathcal{I}^{(6)}_{\{5,1\}}$. Particularly, the tensor products
at general $q$, $V_3 \otimes V_2=V_4\oplus V_2$ and $V_3\otimes
V_3=V_5\oplus V_3\oplus V_1$, degenerate and turn correspondingly
into $\mathcal{I}^{(6)}_{\{4,2\}}$ and
$\mathcal{I}^{(6)}_{\{5,1\}}\oplus V_3$ at $q^{3}=\pm 1$.

The simplest cases for which we can try to find the solutions
correspond to the matrices $\check{R}^{3\; 3}(u)$ and
$\check{R}^{3\; 2}(u)/\check{R}^{2\; 3}(u)$. The spectral
parameter dependent solution  $\check{R}^{2\; 3}(u)$ to the YBE
$(\check{R}^{2\; 2}\check{R}^{2\; 3}\check{R}^{2\;
3}=\check{R}^{2\; 3}\check{R}^{2\; 3}\check{R}^{2\; 2})$ 
  at general $q$ is unique (\ref{rgi}), which is fixed by the
fundamental matrix $\check{R}^{2\; 2}(u)$. If to take  as
$\check{R}^{2\; 2}(u)$ the unity matrix or any other
$sl_q(2)$-symmetric $4\times 4$ matrix, then the solution
$\check{R}^{2\; 3}(u)$ is constant. The same is valid
 at $q^{3}=\pm 1$ as well, when the decomposition
$\check{R}^{2\; 3}(u)=\breve{P}_4+f(u)\breve{P}_2$ smoothly
transforms into
$\breve{P}_{\mathcal{I}^{(6)}_{\{4,2\}}}+\bar{f}(u){\breve{P}}'_{\mathcal{I}^{(6)}_{\{4,2\}}}$
 (see the analysis in the previous section). Here
$\breve{P}_{\mathcal{I}^{(6)}_{\{4,2\}}}=\mathbb{I}$,
${\breve{P}}'_{\mathcal{I}^{(6)}_{\{4,2\}}}=\lim_{q\to e^{i\frac{
r\pi}{3}}}{(c_4-c_2)\breve{P}_2}$ and $\bar{f}(u)=\lim_{q\to
e^{i\frac{ r\pi}{3}}}{(f(u)-1)/(c_4-c_2)}$, $r=1,2,4,5$.

Similarly we must take $\check{R}^{3\;3}(u)$ at $q^{3}=\pm 1$ in
the form of $\check{R}^{3\;3}(u)=P_{\mathcal{I}^{(6)}_{\{5,1\}}}+
f(u)P'_{\mathcal{I}^{(6)}_{\{5,1\}}}+g(u)P_3$. The Casimir
operator on the space of the tensor product $V_3\otimes V_3$  can
be expressed as $c^{3\;
3}=\frac{-1}{3}P_{\mathcal{I}^{(6)}}+P'_{\mathcal{I}^{(6)}}+\frac{2}{3}
P_3$, and $P_{\mathcal{I}^{(6)}}+P_3=\mathbb{I}$. The projectors
$P_5$ and $P_1$ have poles at $q^3=\pm 1$, 
but the solutions (\ref{3solr}, \ref{ra}) are well defined and are
transformed into the following expressions (we have fixed below
$q=(-1)^{1/3}=e^{i \pi/3}$)
\bea \check{R}_{1}^{33}(u)\! =\!
P_{\!\!\!\mathcal{I}^{(6)}_{\{5,1\}}}\!\!\!+\!
\frac{i\sqrt{3}(q^{2u}\!\!-\!1)}{1\!+\!q^u\!+\!q^{2u}}
{P'}_{\!\!\!\mathcal{I}^{(6)}_{\{5,1\}}}\!\!\!+\!\frac{q^{u\!\!+\!1}\!
+\! 1}{q+q^{u}}P_3,\;\;\;\check{R}_2^{33}(u)\!
=\!P_{\!\!\!\mathcal{I}^{(6)}_{\{5,1\}}}\!\!\!
+\!\frac{i\sqrt{3}(q^u\!-\!1)}{1\!+\!q^u}
{P'}_{\!\!\!\mathcal{I}^{(6)}_{\{5,1\}}}\!\!\!\!\!+\!\frac{q^{u\!+\!1}\!\!+\!1}{q+
q^{u}}P_3,\label{r33}
 \eea
$\check{R}^{33}_{\pm}=\mathbb{I}\pm\frac{i(q^u-1)}{1+q^u}
{P'}_{\!\!\!\mathcal{I}^{(6)}_{\{5,1\}}}.\;\;\;$
 There are not new
constant or spectral parameter dependent solutions at roots of
unity also for the YBE with $\check{R}^{32}(u)$ matrix
($\check{R}^{33}\check{R}^{32}\check{R}^{32}=\check{R}^{32}\check{R}^{32}\check{R}^{33}$).
The only spectral parameter solutions are the limit cases of the
corresponding solutions (\ref{rgi}). If to take  in the YBE as
$\check{R}^{33}(u)$ any other $sl_q(2)$-invariant $9\times 9$
matrix, the $\check{R}^{32}(u)$-matrix becomes constant
(equivalent to the constant solution $\check{R}^{23}(u)$).

\paragraph{The solution at $q^6=-1$.} Note, that all of the spectral parameter dependent
solutions discussed up to now are supplemented by the
normalization condition $\check{R}(0)=\mathbb{I}$. We would like
to mention a peculiarity which is met at
$q^6=-1$ ($t^3=1$ for $osp_t(1|2)$ \cite{SHKKH}). %
Here there is no degeneration in the fusion for the tensor product
$V_3\otimes V_3$, but the following solution to the YBE
\cite{SHKKH}
\bea q^6=-1,\quad \check{R}_{o}^{3\; 3}(u) = P_5+\frac{q^4 q^u -
1}{q^4 - q^u}P_3- P_1.\eea
has the property $\check{R}_{o}^{3\; 3}(0)=P_5+P_3-P_1$. At first
sight this solution coincides with the solution $\check{R}^{3\;
3}_{2}(u)$ in (\ref{3solr}), if to take the limit $q\to
(-1)^{r/6},\; r=1,3,5,7,9,11$. But there is a notable difference
at the point $u=0$, where both of $\check{R}^{3\; 3}_{1,2}(0)$
(\ref{3solr}) become unity matrices, which is important. It means,
that $\lim_{q\to (-1)^{r/6}}\lim_{u\to 0}\check{R}_{2}^{3\;
3}(u)\neq\lim_{u\to 0}\lim_{q\to (-1)^{r/6}}\check{R}_{2}^{3\;
3}(u)$.  Note, that for  $q^4=1$ the matrix $\check{R}_{o}$ is a
solution too (and the peculiarities noted above about the
not-coinciding limits are right also here), but as we know for
this case $V_3$ is not an irrep. We can denote it as a
$\bar{V}_3\supset V_1$ (as in \cite{kkh}) and write the proper
fusion $\bar{V}_3\otimes
\bar{V}_3=\mathcal{I}^{(8)}_{\{5,3\}}\oplus V_1$, where
$\mathcal{I}^{(8)}_{\{5,3\}}$ is equivalent to the direct sum of
two $\mathcal{I}^{(4)}_{\{3,1\}}$. We shall not analyze this case,
as it is included in a non-direct way in 
consideration of $\bigotimes^4
V_2=\mathcal{I}^{(4)}_{\{3,1\}}\otimes
\mathcal{I}^{(4)}_{\{3,1\}}$ (as
$\mathcal{I}^{(4)}_{\{3,1\}}\supset{\bar{V}_3}$ (\cite{kkh})) done
further in this section.

\paragraph{Some notes and statements.} The expressions
above (\ref{r33}) can be obtained either by direct solving of the
YBE at roots of unity or by taking the corresponding limits of the
solutions existing at general $q$, using appropriate modifications
of the expressions. When at $q^{n}=1$ in the fusion of two irreps
indecomposable representation
$\mathcal{I}^{(\mathcal{R})}_{\{r,\mathcal{R}-r\}}$ 
arises from the merging of the representations $V_{r}$ and
$V_{\mathcal{R}-r}$, and the projectors $P_{\mathcal{R}-r}$ and
$P_{r}$ acquire singularities \cite{kkh}, the Casimir operator
remains well defined and  can be rewritten in terms of the
projectors $P_{\mathcal{I}^{(\mathcal{R})}_{\{r,\mathcal{R}-r\}}}$
and $P'_{\mathcal{I}^{(\mathcal{R})}_{\{r,\mathcal{R}-r\}}}$. As
at general $q$ the projectors $P_{\mathcal{R}-r}$ and $P_{r}$ are
included in $c$ as the sum $c_{\mathcal{R}-r}P_{\mathcal{R}-r}+c_r
P_{r}$, we can rewrite it as $c_r
(P_{r}+P_{\mathcal{R}-r})+(c_{\mathcal{R}-r}-c_r)
P_{\mathcal{R}-r}$, where the first summand
$P_{r}+P_{\mathcal{R}-r}$ transforms at roots of unity to the
projector $P_{\mathcal{I}^{(\mathcal{R})}_{\{r,\mathcal{R}-r\}}}$
and the second one to the projector $(c_{\mathcal{R}-r}-c_r)/c_r
P_{\mathcal{R}-r}\Rightarrow
P'_{\mathcal{I}^{(\mathcal{R})}_{\{r,\mathcal{R}-r\}}}$. At the
given roots of unity 
the Casimir operator becomes degenerate, $c_{\mathcal{R}-r}=c_r$,
and here the singularity in the projector $P_{\mathcal{R}-r}$ has
been canceled by the zero in the nominator. Putting in the
expression of the matrix $\check{R}_{VV}(u)$ the projectors
$P_{\mathcal{R}-r}$ and $P_{r}$ written in terms of
$P_{\mathcal{I}^{(\mathcal{R})}_{\{r,\mathcal{R}-r\}}}$ and
$P'_{\mathcal{I}^{(\mathcal{R})}_{\{r,\mathcal{R}-r\}}}$, and then
taking the corresponding values of $q$ we shall obtain the exact
well-defined expression. 
This is
conditioned by the fact, that the coefficients of the projectors
$P_{\mathcal{R}-r}$ and $P_{r}$ in the expansion of
$\check{R}_{VV}(u)$ (\ref{rij}) coincide at the corresponding
roots of unity, as it was for the case of the Casimir operator.

Essentially new solutions to the YBE can be obtained in the cases,
when the number of the projectors at roots of unity increases
comparing with the case of general $q$. It happens when we 
consider matrices $\check{R}_{V\mathcal{I}}$ and
$\check{R}_{\mathcal{I}\mathcal{I}}$ acting on the tensor products
$V_r\otimes \mathcal{I}^{(\mathcal{R})}_{\{r',\mathcal{R}-r'\}}$
and $\mathcal{I}^{(\mathcal{R})}_{\{r,\mathcal{R}-r\}}\otimes
\mathcal{I}^{(\mathcal{R}')}_{\{r',\mathcal{R}'-r'\}}$, which
stand instead of $V_r\otimes(V_{r'}\oplus V_{\mathcal{R}'-r'})$
and $(V_{r}\oplus V_{\mathcal{R}-r})\otimes(V_{r'}\oplus
V_{\mathcal{R}'-r'})$ at general $q$.  We shall analyse the 
simplest such case below, when $q=i$. We can calculate that the
number of the linear independent $r\mathcal{R} \times r\mathcal{R}
$- and $\mathcal{R}^2\times \mathcal{R}^2$-matrices (hence, the
number of the independent projectors also) acting on the
$r\mathcal{R}$ and $\mathcal{R}^2$-dimensional representation
spaces of the mentioned tensor products at general $q$ and at
roots of unity ($q^{\mathcal{R}}$=1) are different.
 Hereafter we
shell refer as new solutions (providing $q$ is a root of unity) to
those, which are not obtained  at roots of unity from the
solutions existing at general $q$.

\subsection{YBE solutions at $q=i$.}
At $q^4=1$ (we fix $q=i$, the case of $q=-i$ is completely
equivalent to this case) only two non-reducible highest weight
representations exist in the fusions of the fundamental
two-dimensional spin-$1/2$ representations. They are
two-dimensional spin-$1/2$ irrep $V_2$ and four-dimensional
indecomposable representation $\mathcal{I}^{(4)}_{\{3,1\}}=V_2
\otimes V_2$. The tensor product decomposition rules for them have
the following form
\bea \otimes^2 V_2=\mathcal{I}^{(4)}_{\{3,1\}},\; \; \; \; \;
V_2\otimes \mathcal{I}^{(4)}_{\{3,1\}}=\oplus^4 V_2,\; \; \; \; \;
\otimes^2 \mathcal{I}^{(4)}_{\{3,1\}}=\oplus^4
\mathcal{I}^{(4)}_{\{3,1\}}.\label{vii}\eea

The corresponding YBE for the matrices ${R}^{2\; 2}$,
$\check{R}^{2\; 4}$ and $\check{R}^{4\; 4}$ are
\bea\label{r22}\Big(\check{R}^{2\; 2}(u)\otimes
\mathbb{I}\Big)\!\Big(\mathbb{I}\otimes \check{R}^{2\; 2}(u+
v)\!\Big)\!\Big( \check{R}^{2\;
2}(v)\otimes\mathbb{I}\Big)\!=\!\Big(\mathbb{I}\otimes\check{R}^{2\;
2}(v)\!\Big)\!\Big(\check{R}^{2\; 2}(u+
v)\otimes\mathbb{I}\Big)\!\Big(\mathbb{I}\otimes\check{R}^{2\;
2}(v)\!\Big),\\ \label{r24} \Big(\check{R}^{2\; 2}(u)\otimes
\mathbb{I}\Big)\!\Big(\mathbb{I}\otimes \check{R}^{2\; 4}(u+
v)\!\Big)\!\Big( \check{R}^{2\;
4}(v)\otimes\mathbb{I}\Big)\!=\!\Big(\mathbb{I}\otimes\check{R}^{2\;
4}(v)\!\Big)\!\Big(\check{R}^{2\; 4}(u+
v)\otimes\mathbb{I}\Big)\!\Big(\mathbb{I}\otimes\check{R}^{2\;
2}(v)\!\Big),\\
\Big(\check{R}^{4\; 4}(u)\otimes
\mathbb{I}\Big)\!\Big(\mathbb{I}\otimes \check{R}^{4\; 4}(u+
v)\!\Big)\!\Big(\check{R}^{4\; 4}(v)\otimes
\mathbb{I}\Big)\!=\!\Big(\mathbb{I}\otimes\check{R}^{4\;
4}(v)\!\Big)\!\Big(\check{R}^{4\; 4}(u+ v)\otimes
\mathbb{I}\Big)\!\Big(\mathbb{I}\otimes \check{R}^{4\;
4}(u)\!\Big),\label{r44}\eea
acting accordingly on the vector spaces $V_2\otimes V_2\otimes
V_2$, $V_2\otimes V_2\otimes \mathcal{I}^{(4)}_{\{3,1\}}$ and
$\mathcal{I}^{(4)}_{\{3,1\}}\otimes
\mathcal{I}^{(4)}_{\{3,1\}}\otimes\mathcal{I}^{(4)}_{\{3,1\}}$.
Here we have preferred to write the action of the operators in the
tensor product form to avoid the usual lower indexes (see e.g. Eq.
(\ref{rrr})), which distinguish different spaces, meanwhile the
indexes used here denote the dimensions of the representation
spaces.

Note, that also the YBE defined on the space
$\mathcal{I}^{(4)}_{\{3,1\}}\otimes
\mathcal{I}^{(4)}_{\{3,1\}}\otimes V_2$ could be considered,
\bea  \label{r42} \Big(\check{R}^{4\; 4}(u)\otimes
\mathbb{I}\Big)\!\Big(\mathbb{I}\otimes \check{R}^{4\; 2}(u+
v)\!\Big)\!\Big( \check{R}^{4\;
2}(v)\otimes\mathbb{I}\Big)\!=\!\Big(\mathbb{I}\otimes\check{R}^{4\;
2}(v)\!\Big)\!\Big(\check{R}^{4\; 2}(u+
v)\otimes\mathbb{I}\Big)\!\Big(\mathbb{I}\otimes\check{R}^{4\;
4}(v)\!\Big),\eea
the solutions of which are not necessarily the solutions to the
equations (\ref{r24}) and (\ref{r44}). Here we shall concentrate
on the YBE (\ref{r24}) and (\ref{r44}).

There is a unique non-trivial solution ${R}^{2\; 2}(u)$ to
(\ref{r22}), which is just the limit  $q\to i$ of the solution
(\ref{rgi}), $\check{R}^{2\;
2}(u)=\mathbb{I}+\frac{i(1-\mathrm{e}^u)}{1+\mathrm{e}^u}c^{2\;
2}$ (we have chosen the parametrization taking into account the
freedom of the normalization of the spectral parameter, to replace
$q^u$ with $\exp{(u)}$, which is a convenient expression for the
fixed values of $q$). ${R}^{2\; 2}(u)$ can be expressed also by
means of two projection operators,
$P_{\mathcal{I}^{(4)}_{\{3,1\}}}$($=\mathbf{I}$) and
$P'_{\mathcal{I}^{(4)}_{\{3,1\}}}$ ($\approx\lim_{q\to
i}{(c_3-c_1)P_1}$).

\subsubsection{The solutions $\check{R}^{2\; 4}(u)$.}

  The two-dimensional spaces in the decomposition of
$V_2\otimes \mathcal{I}^{(4)}_{\{3,1\}}$ (\ref{vii}) must be
considered pairwise, ${\tilde{V}}_2^i,\; i=1,2$ (two
representations, emerging from the splitting of the representation
$V_4$ in $\bigotimes^3 V_2$ at $q=i$) and the remaining two
${V}_2^i,\; i=1,2$: $V_2\otimes V_2\otimes V_2=V_4\oplus V_2\oplus
V_2\Rightarrow_{q\to i} \tilde{V}_2\oplus \tilde{V}_2\oplus V_2
\oplus V_2 $,  as they have Casimir eigenvalues $c_4,\; c_2$
differing by sign at $q=i$. Thus the projection operators now are
eight, ${\tilde{P}}_2^{ij}$ and $P_2^{ij},\; i,j=1,2$ (at general
$q$ they are five, $P_4$ and $P_2^{ij},\; i,j=1,2$). As here we
have larger space of the projectors than for the case of general
$q$, we can look for new solutions 
in the form
\bea
R^{2\;4}(u)=\sum_{i,j=1,2}\Big({\tilde{f}}_{ij}(u){\tilde{P}}_2^{ij}+f_{ij}(u)P_2^{ij}
\Big).\label{rf24} \eea

 Taking in the YBE (\ref{r24}) the intertwiner
$\check{R}^{22}(u)=\mathbb{I}+\frac{i(1-\mathrm{e}^u)}{1+\mathrm{e}^u}c^{2\;
2}$, we find that the only spectral parameter dependent solution
of $\check{R}^{2\; 4}(u)$ with the normalization property
$\check{R}^{24}(0)=\mathbb{I}$, is given as follows
\bea \label{ri24}\check{R}^{2\; 4}(u)\!=\!
\Big[{\tilde{P}}_2^{11}\!+{\tilde{P}}_2^{22}\Big]\!+\!\frac{1+6
\mathrm{e}^u\!+\mathrm{e}^{2u}}{2(1+\mathrm{e}^u)^2}\Big[P_2^{11}\!+P_2^{22}\Big]\!
+\!\frac{i(\mathrm{e}^u-1)}{2(1+\mathrm{e}^u)^2}\Big[P_2^{12} (1+3
\mathrm{e}^u)\!+P_2^{21}(3+\mathrm{e}^u)\Big]\!.\eea
 This matrix corresponds to the ordinary XX model.
 It is just the composite solution 
 $\check{R}^{2\; 4}(u)=\Big(R^{22}(u)\otimes \mathbf{I}\Big)\Big(\mathbf{I}\otimes R^{22}(u)\Big)$
at $q=i$. Such solution could be obtained also from Jimbo's
composite relations (\ref{jc}, \ref{jimbo_comp}). The relation
(\ref{jimbo_comp1}) provides with the solution
$\Big(R^{22}(u)\otimes \mathbf{I}\Big)\Big(\mathbf{I}\otimes
R^{22}(u+u_0)\Big)$  at general $q$ and in the limit $q\to i$,
too. At $q=i$ there is also another generalization of the matrix
(\ref{ri24}), for which $R^{24}(0)\neq \mathbb{I}$, and where the
projectors ${\tilde{P}}_2^{11}$ and ${\tilde{P}}_2^{22}$ have
different coefficient functions containing an arbitrary parameter
$f_0$. This means that such solution could not exist at general
$q$, as in the limit $q\to i$ the projectors ${\tilde{P}}_2^{ij}$
appear only in the following sum,
$P_4\Rightarrow{\tilde{P}}_2^{11}\!+{\tilde{P}}_2^{22}$. The
general expression of that solution is the following
\bea \label{rgf24}&\check{R}^{2\; 4}(u;u_0,f_0)\!=\!
2\Big((1+f_0)(1+\cosh{[u_0]})+\cosh{[u]}+\cosh{[u+u_0]}+(1-f_0)\sinh{[u_0]}\Big)
{\tilde{P}}_2^{11}\!+&\nn\\\nn &
2\Big((1-f_0)(1+\cosh{[u_0]})+\cosh{[u]}+\cosh{[u+u_0]}
+(1+f_0)\sinh{[u_0]}\Big){\tilde{P}}_2^{22}\!+&\\&
\!\Big(4-f_0+\cosh{[u]}+(2-3 f_0)\cosh{[u_0]}+\cosh{[u+u_0]}+3 f_0
\sinh{[u_0]}\Big)P_2^{11}\!+&\\\nn & \Big(4+3f_0+\cosh{[u]}+(2+
f_0)\cosh{[u_0]}+\cosh{[u+u_0]}- f_0 \sinh{[u_0]}\Big)P_2^{22}\!
+&\\\nn & \! i\Big(f_0+\cosh{[u]}-(2+
f_0)\cosh{[u_0]}+\cosh{[u+u_0]}+ f_0 \sinh{[u_0]}+2
\sinh{[u+u_0]}+2 \sinh{[u]}\Big) P_2^{12} \!-&\\\nn &
i\Big(f_0+\cosh{[u]}-(2+ f_0)\cosh{[u_0]}+\cosh{[u+u_0]}+f_0
\sinh{[u_0]}-2 \sinh{[u+u_0]}-2 \sinh{[u]}\Big)P_2^{21}\!. \eea
When $f_0=0$ and $u_0=0$ this expression coincides with the
solution (\ref{ri24}), after multiplying by an overall function.
This expression is a solution to the YBE, and also obeys to
(\ref{jimbo_comp1}), but the generalization for $w=u+u_0$ of the
next composite relation (\ref{jc})  fixes $f_0=0$.

The other spectral parameter dependent solution,
 which exists at general $q$ is the representation
of the matrix $\check{R}^{2\;3}(u)$ in the space $V_2 \otimes V_2
\otimes V_2$, which we shall denote as
$\check{R}^{2\;4_{(3)}}(u)$. This is the solution of  Jimbo's
ordinary relation (\ref{jimbo}). This solution also contains an
arbitrariness coming from the combination of the projectors $\sum'
P^{ij}_2$($=q/\sqrt{1+q^2+q^4} P^{11}_2-\sqrt{1+q^2+q^4}/q
P^{22}_2+q^2 P^{12}_2-q^{-2}P^{21}_2$ in the basis fixed by us)
which vanishes after multiplication by the operators containing
$P_3$, $[P_3\otimes \mathbb{I}]\sum' P^{ij}_2[\mathbb{I}\otimes
P_3]=0$. Actually it is proportional to the matrix $[P_1\otimes
\mathbb{I}][\mathbb{I}\otimes P_1]$. Here $P_1$ and $P_3$ are the
$4\times 4$ projector operators into the one- and three-
dimensional spaces in the fusion at general $q$ ($V_2\otimes
V_2=V_1\oplus V_3$), $\mathbb{I}$ is the $2\times 2$ unity matrix.
A term $f(u)\sum' P^{ij}_2$ with arbitrary coefficient function
$f(u)$ can be added to $\check{R}^{2\;4_{(3)}}(u)$, and it will
remain as a solution to the YBE or  Jimbo's ordinary relation (at
any values of $q$). We learn also from these examples, that the
existence of the arbitrary functions in the solutions can speak
about the possibility to reduce the YBE on the subspaces of the
representations (for the given example two separate parts of the
matrix $R^{24}(u)$ are acting separately on the subspace $V_2
\otimes V_3$ and the subspace $V_2 \otimes V_1$ of the entire
space $V_2\otimes V_2 \otimes V_2=V_2\otimes(V_3\oplus V_1)$.)

 At $q=i$ this solution contains a singularity, and if to take the
limit $q\to i$ after multiplying by $(1+q^2)$, the solution
becomes constant one. One can note that the singular term is
proportional to the matrix $\sum' P^{ij}_2$, so by adding to this
solution a matrix $\sum' P^{ij}_2$ with appropriate defined
coefficient function, we can remove the singularity and have a
good defined limit  $q\to i$ (below $f(u)$ is an arbitrary
function  and $P^{11}_2+P^{22}_2+i P^{12}_2-i P^{21}_2=\sum'
P^{ij}_2$)
\bea
\check{R}^{2\;4_{(3)}}(u)=({\tilde{P}}_2^{11}\!+{\tilde{P}}_2^{22})+i\frac{1+e^{u+u_0}}{e^{u+u_0}-1}(P_2^{12}+P_2^{21})+
f(u)(P^{11}_2+P^{22}_2+i P^{12}_2-i P^{21}_2).\label{rg24}
 \eea
Here $u_0$ is  an arbitrary number: the shifting of the spectral
parameter is a permissible transformation of the solutions.

We see, as it was expected, that the consideration of Jimbo's
relations gives only particular solutions, 
so in the following we shall deal straightforwardly with the YBE
(\ref{r24}) and (\ref{r44}).

  There are numerous constant solutions to (\ref{r24}) at $q=i$. Some of them are the limit cases
  of the 
  spectral parameter dependent solutions taken at $u\to 0, \pm
  \infty$.
  We would like to present below only such solutions, which could be considered
  as new ones (with existence of ${\tilde{P}}_2^{ij}$ projectors with different
  coefficients).
   Such
 constant solutions $\check{R}^{2\; 4}_c(u)$ are 
\bea \nn&\check{R}^{2\; 4}_c(u)={\tilde{P}}_2^{22}+g_0
P_2^{11}+\frac{g_0-2}{2 g_0-1}\left({g_0
\tilde{P}}_2^{11}+P_2^{22}\right),&\\ &\check{R}^{2\;
4}_c(u)={\tilde{P}}_2^{22}+\frac{g_0^2-2(f_0+g_0)}{2(\!f_0+g_0\!)-1}{\tilde{P}}_2^{11}
+g_0 {P}_2^{11}+\frac{g_0-2(f_0
g_0+2)}{2(\!f_0+g_0\!)-1}{P}_2^{22}\!+\!f_0
\left(\!P^{11}_2+P^{22}_2\!+\!i P^{12}_2\!-\!i
P^{21}_2\!\right),\;\;\;&\\
&\check{R}^{2\;
4}_c(u)=g_0\Big({\tilde{P}}_2^{11}-{\tilde{P}}_2^{22}+{\tilde{P}}_2^{22}-
P_2^{11}\Big)+f_0 \left(P^{11}_2+P^{22}_2+i P^{12}_2-i
P^{21}_2\right).&\nn\eea
Here $g_0$ and $f_0$ are arbitrary constants. And, moreover, all
these matrices satisfy to the YBE (\ref{r24})  with arbitrary
$sl_i(2)$ invariant $\check{R}^{2\;2}(u)$, i.e.
$\check{R}^{2\;2}(u)=\mathbb{I}+\mathrm{f}(u) c^{2\; 2}$, where
$\mathrm{f}(u)$ can be any function.
 Spectral parameter dependent solutions with the arbitrary
$R^{2\;2}(u)$-matrix also exist (with ${\tilde{P}}_2^{12}$ or $
{\tilde{P}}_2^{21}$),
\bea R^{2\;4}(u)= {\tilde{P}}_2^{12/21}+ g(u)(P^{11}_2+P^{22}_2+i
P^{12}_2-i P^{21}_2).\eea
The second part of this solution with arbitrary function $g(u)$ is
a constant solution also at general $q$ (i.e. the matrix
$\sum'P^{ij}_2$).

 Also we would like to
mention the following two solutions,
\bea \check{R}^{2\;
4}=\tilde{f}_1(u)\tilde{P}_2^{11}+\tilde{f}_2(u)\tilde{P}_2^{22}+f(u)(P^{11}_2+P^{22}_2+i
P^{12}_2-i P^{21}_2)\eea
 and
\bea &\check{R}^{2\;
4}(u)=h(u)\Big(\sum_{i,j=1,2}\tilde{h}_{ij}{\tilde{P}}_2^{ij}+\Big[P_2^{12}+P_2^{21}\Big]\Big)+
f(u)(P^{11}_2+P^{22}_2+i P^{12}_2-i P^{21}_2).&\eea
(with arbitrary functions $\tilde{f}_{1,2}(u),\; f(u)$ and $h(u)$
and arbitrary numbers $\tilde{h}_{ij}$) which satisfy to the YBE
with $4\times 4$ intertwiner matrix $R^{22}(u)=\mathbb{I}$. It
means, that together with the transfer matrices with different
spectral parameters, constructed via the given $R$-matrices,  the
monodromy matrices also are commuting. As there is no proper
normalization for both matrices to give
$\check{R}(u_0)=\mathbb{I}$ at some point $u_0$, so we shall not
try to investigate the chain models corresponding to such
matrices.

\subsubsection{The solutions $\check{R}^{4\; 4}(u)$.}

According to (\ref{vii}) the decomposition $\otimes^2
\mathcal{I}^{(4)}_{\{3,1\}}$ contains four
$\mathcal{I}^{(4)}_{\{3,1\}}$-representations. One must note here,
that although all $\mathcal{I}^{(4)}_{\{3,1\}}$ are isomorphic one
to another, they have different sets of the eigenvalues of the
$k$-operator. Schematically one can describe the representation
$\mathcal{I}^{(4)}_{\{3,1\}}=\{v_+,v_0,v_-,u_0\}$ as follows
\bea &e\cdot \{v_+,v_0,v_-,u_0\}=\{0,0,v_0,v_+\},&\nn\\&f\cdot
\{v_+,v_0,v_-,u_0\}=\{v_0,0,0,v_-\},&\nn\\&k\cdot
\{v_+,v_0,v_-,u_0\}=\varepsilon \{v_+,-v_0,v_-,-u_0\},&
\label{iefkc}\\& c\cdot \{v_+,v_0,v_-,u_0\}=\{0,0,0,v_0\}.&\nn\eea
Some numerical coefficients' variation is possible in this
schematic action, due to the normalization of the vectors. The
sign $\varepsilon=\pm$ is positive for two representations and is
negative for the other pair.
 This happens from the following
reason. The fusion of the tensor product $V_2\otimes V_2\otimes
V_2 \otimes
 V_2$ at general $q$ is
 $V_5\oplus \Big[\bigoplus_{i=1}^3 V_3^i\Big]\oplus \Big[\bigoplus_{i=1}^2
 V_1^i\Big]$. At $q=i$ two three-dimensional and two one-dimensional
 representations deform into two indecomposable ones, $V_3\oplus V_1 \Rightarrow
\mathcal{I}^{(4)}_{\{3,1\}}$, with $\varepsilon=-$. Meanwhile the
other two indecomposable representations emerge from the
deformation and splitting to the direct sum in this way $V_5\oplus
V_3\Rightarrow \mathcal{I}^{(8)}_{\{5,3\}}\Rightarrow
\mathcal{I}^{(4)}_{\{3,1\}}\oplus \mathcal{I}^{(4)}_{\{3,1\}}$
(see the work \cite{kkh} for details), with $\varepsilon=+$.

 Let us denote four
indecomposable representations by
$\mathcal{I}^{(4)i}_{\{3,1\}\pm}=\{v_+,v_0,v_-,u_0\}_{\pm}^{i}$,
$i=1,2$. The possible independent projectors are
$P^{ij}_{\mathcal{I}\;\varepsilon \eta}$,
${P'}^{ij}_{\mathcal{I}\;\varepsilon \eta}$, where
$\varepsilon,\eta\in\{+,-\}$ and $i,j\in\{1,2\}$. The action of
the projectors $P^{ij}_{\mathcal{I}\;\varepsilon \varepsilon}$,
${P'}^{ij}_{\mathcal{I}\;\varepsilon \varepsilon}$ corresponds to
the description given in the previous sections,
\bea P^{ij}_{\mathcal{I}\;\varepsilon \varepsilon}\cdot
 \{v_+,v_0,v_-,u_0\}_{\varepsilon}^{j}=\{v_+,v_0,v_-,u_0\}_{\varepsilon}^{i},\\
{P'}^{ij}_{\mathcal{I}\;\varepsilon \varepsilon}\cdot
 \{v_+,v_0,v_-,u_0\}_{\varepsilon}^{j}=\{0,0,0,v_0\}_{\varepsilon}^{i}. \eea
Meanwhile, the action of the projectors
$P^{ij}_{\mathcal{I}\;\varepsilon \; \bar{\varepsilon}}$,
${P'}^{ij}_{\mathcal{I}\;\varepsilon\; \bar{\varepsilon}}$, where
$\bar{\varepsilon}$ is the opposite sign of $\varepsilon$, can be
defined in the following way,
\bea P^{ij}_{\mathcal{I}\;\varepsilon \; \bar{\varepsilon}}\cdot
 \{v_+,v_0,v_-,u_0\}_{\bar{\varepsilon}}^{j}=\{v_0,0,0,v_-\}_{\varepsilon}^{i},\\
{P'}^{ij}_{\mathcal{I}\;\varepsilon \; \bar{\varepsilon}}\cdot
 \{v_+,v_0,v_-,u_0\}_{\bar{\varepsilon}}^{j}=\{0,0,v_0,v_+\}_{\varepsilon}^{i}.
\eea
In summary there are $32$ independent projectors or algebra
invariants (in explicit form they are given in the Appendix) in
the representation space $\bigotimes^4 V_2=\bigotimes^2
\mathcal{I}^{(4)}_{\{3,1\}}$ and hence the $R$-matrix can be
constructed by means of their sum with $32$ coefficient functions
(one of them can be chosen as $1$ due to
 normalization freedom).
  At general $q$ the number of the independent projectors is
$14$: $P_5,\;P_{3}^{ij}$ and $P_1^{kr}$ with $i,j=1,2,3$ and
$k,r=1,2$.

 The simplest solution at general $q$ can be obtained just by the
 following tensor product on the vector space $V_2\otimes V_2\otimes V_2 \otimes
 V_2$, using the fundamental solution $\check{R}^{2\;
2}(u)$ on the spin-$\frac{1}{2}$ states (the descendant property
has been used)
\bea \check{R}^{4\; 4}(u)=\Big(\mathbb{I}\otimes\check{R}^{2\;
2}(u)\otimes\mathbb{I}\Big)\Big(\check{R}^{2\;
2}(u)\otimes\mathbb{I}\otimes\mathbb{I}\Big)\Big(\mathbb{I}\otimes
\mathbb{I}\otimes\check{R}^{2\;
2}(u)\Big)\Big(\mathbb{I}\otimes\check{R}^{2\;
2}(u)\otimes\mathbb{I}\Big).\label{rr4}\eea
Here $\mathbb{I}$ is the $2\times 2$ unity operator defined on the
space $V^2$. This $\check{R}$-matrix can be expressed surely by
the mentioned above $14$ projectors. Some modifications are
possible of this solution conditioned by the automorphisms of the
algebra, but it does not change the nature of the solution. At the
limit $q\to i$ the linear combination of the projectors
$P_5,\;P_{3}^{ij}$ and $P_1^{kr}$ 
in the $R^{4\; 4}$ can be expressed by the sum of the following
fourteen projectors -
$\Big(P^{11}_{\mathcal{I}++}+P^{22}_{\mathcal{I}++}\Big),\;
P^{11}_{\mathcal{I}--},\;P^{22}_{\mathcal{I--}},
\;P^{12}_{\mathcal{I}--},\;P^{21}_{\mathcal{I}--},\;
\Big(P'^{11}_{\mathcal{I}++}+P'^{22}_{\mathcal{I}++}\Big),\;
P'^{11}_{\mathcal{I}--},\;P'^{22}_{\mathcal{I--}},\;
P'^{12}_{\mathcal{I}--},\; P'^{21}_{\mathcal{I}--},\;
\Big(P'^{11}_{\mathcal{I}-+}-P^{12}_{\mathcal{I}-+}\Big),\;
\Big(P'^{21}_{\mathcal{I}+-}-P^{11}_{\mathcal{I}+-}\Big),\;\Big(P'^{21}_{\mathcal{I}-+}-
P^{22}_{\mathcal{I}-+}\Big),\;\Big(P'^{22}_{\mathcal{I}+-}-
P^{12}_{\mathcal{I}+-}\Big)$, which can be found as the limits
$q\to i$ of the appropriate linear combinations of the projectors
at general $q$. The explicit expression of $ \check{R}^{4\; 4}(u)$
is the following (below
 $t=\tanh{u}$)
\bea \label{r4xx}&\check{R}^{4\;
4}(u)=P^{11}_{\mathcal{I}++}+P^{22}_{\mathcal{I}++}+(1-2
t^2+t^3)P^{11}_{\mathcal{I}--}+(1-2
t^2-t^3)P^{22}_{\mathcal{I--}}+&\\\nn&t(2-t^2)[P^{12}_{\mathcal{I}--}-P^{21}_{\mathcal{I}--}]
+it[P'^{11}_{\mathcal{I}++}+P'^{22}_{\mathcal{I}++}]+\frac{i
}{2}t(-8+t+5t^2-t^3)P'^{11}_{\mathcal{I}--}+&\\\nn&\frac{i}{2}t(4-t-t^2+t^3)P'^{22}_{\mathcal{I--}}+
\frac{i}{2}t(-6-3t+t^3)P'^{12}_{\mathcal{I}--}+\frac{i}{2}t(-6+3
t+6
t^2-t^3)P'^{21}_{\mathcal{I}--}+&\\\nn&t(1-t)(\frac{i}{2}[P'^{11}_{\mathcal{I}-+}-P^{12}_{\mathcal{I}-+}]
+[P^{11}_{\mathcal{I}+-}-P'^{21}_{\mathcal{I}+-}])+t(1+t)(\frac{i}{2}[P^{22}_{\mathcal{I}-+}-P'^{21}_{\mathcal{I}-+}]+
[P^{12}_{\mathcal{I}+-}-P'^{22}_{\mathcal{I}+-}]).&\eea
From the previous example we can expect that at $q=i$ there will
be a generalization of this matrix (as the matrix (\ref{rgf24}))
containing more than the mentioned $14$ projectors, and having no
analogue at general $q$. It can be obtained by using one pair
(containing only the generators $e$ and $k^{\pm 1/2}$ or $f$ and
$k^{\pm 1/2}$) of  Jimbo's composite relations (which now involve
three equations for each of the generators $e$ and $f$)
 derived for the case $R^{r'_1\times r''_1\; r'_2\times r''_2}$.
Simultaneous solution of all the relations will coincide exactly
with (\ref{r4xx}).

A generalization of the solution (\ref{rr4}) which exists at any
$q$ can be written as follows (now with  dependence on three
spectral parameters $u,\;v,\;w$, which leads to  corresponding
modifications in the spectral parameter dependence in the YBE
(\ref{r44}))
\bea &\nn\check{R}^{4\; 4}(u;v,w)=\Big(\check{R}^{2\;
2}(v)\otimes\mathbb{I}\otimes\mathbb{I}\Big)\Big(\mathbb{I}\otimes
\mathbb{I}\otimes\check{R}^{2\;
2}(w)\Big)&\\&\Big(\mathbb{I}\otimes\check{R}^{2\;
2}(u)\otimes\mathbb{I}\Big)\Big(\check{R}^{2\;
2}(u-v)\otimes\mathbb{I}\otimes\mathbb{I}\Big)\Big(\mathbb{I}\otimes
\mathbb{I}\otimes\check{R}^{2\;
2}(u-w)\Big)\Big(\mathbb{I}\otimes\check{R}^{2\;
2}(u-v-w)\otimes\mathbb{I}\Big).\label{rru4}&\eea
The matrix (\ref{rr4}) is the particular case of the expression
(\ref{rru4}) with the parameters $w=0=v$, note that
$\check{R}^{2\; 2}(0)=\mathbb{I}$. The matrix representation of
$\check{R}^{3\; 3}(u)$ in $4\times 4$-dimensional representation
space equals to $\check{R}^{4\; 4}(u;1,1)$, as $\check{R}^{2\;
2}(1)=P_3$. This solution can be obtained also from  Jimbo's
ordinary relations (\ref{jimbo}). As in the previous case, 
this solution also admits adding to it some terms ($\approx
P_1\otimes P_1$) with arbitrary coefficient functions, vanishing
after multiplication by $P_3\otimes P_3$. The limit $q\to i$ can
be evaluated as in the case of $R^{2\;4_{(3)}}(u)$.  But we shall
not explicitly consider it now, as well as the generalization of
the solution (\ref{r4xx}), because we are interested in such (new)
solutions which have the normalization property
$\check{R}^{4\;4}(u_0)=\mathbb{I}$ at some $u_0$.

The increasing of the number of the independent projectors from
$14$ to $32$ at $q=i$ allows to hope, that  for the
$\check{R}^{4\; 4}(u)$-matrix besides of the solutions  at general
$q$ there must be also new solutions to the YBE (\ref{r44}).

 As we are interested in the
solutions to the YBE at roots of unity, let us consider the
$\check{R}^{4\; 4}$-matrix in the form of the following linear
expansion over all $32$  projection operators
\bea \check{R}^{4\; 4}(u)=\sum_{i,j,k=1}^2 \Big( f^{ij}_{k}(u)
P^{ij}_{\mathcal{I}\;\varepsilon_k
\varepsilon_k}+{f'}^{ij}_{k}(u){P'}^{ij}_{\mathcal{I}\;\varepsilon_k
\varepsilon_k}+\bar{f}^{ij}_{k}(u)P^{ij}_{\mathcal{I}\;\varepsilon_k
\;
\bar{\varepsilon}_k}+{{\bar{f'}}}^{ij}_k(u){P'}^{ij}_{\mathcal{I}\;\varepsilon_k
\; \bar{\varepsilon}_k}\Big). \label{r3244}\eea

Below we present a list of the spectral parameter dependent
solutions  for the particular cases (if the condition
$\check{R}^{4\; 4}(0)=\mathbb{I}$ fulfills, then the full list for
each case), when some functions in the expansion (\ref{r3244}) are
vanishing.

\paragraph{1.} At the first let us look for a solution in the form
of $\check{R}(u)= a \mathbb{I}+\sum_{ij\;\varepsilon}
f^{ij}_{\varepsilon}(u)
P'^{ij}_{\mathcal{I}\varepsilon\varepsilon}$. When $i=j$ we find
one solution with few arbitrary parameters $f_0^k$,
\bea \check{R}(u)=\mathbb{I}+u (f_0^1
P'^{11}_{\mathcal{I}--}+f_0^2 P'^{22}_{\mathcal{I}--} +f_0^3
P'^{11}_{\mathcal{I}++}+f_0^3 P'^{22}_{\mathcal{I}++}).
\label{rcf} \eea
When $f_0^1=f_0^2=f_0^3=f_0$ then
$\check{R}(u)=\mathbb{I}+u f_0 c^{2222}$, where $c^{2222}$ is the
representation of the Casimir operator $c$ (\ref{cas}) on the
space $V_2\otimes V_2\otimes V_2\otimes V_2$. Note that the
 $c$-operator writes as a sum of 
 the following
four projectors: $P'^{ii}_{\mathcal{I}\varepsilon\varepsilon},\;
i=1,2,\;\varepsilon=\pm$, as the eigenvalues of the $c$-operator
on the eigenvectors $\{v_+,v_0,v_-\}_{\varepsilon}^{i}$ are $0$.

The solutions, when $i\neq j$ in the sum $\sum_{ij\;\varepsilon}
f^{ij}_{\varepsilon}(u)
P'^{ij}_{\mathcal{I}\varepsilon\varepsilon}$, are numerous. Here
we are presenting almost the full list of them, some constant
solutions may have been omitted  (the numbers
$f_0,\;g_0,\;h_0,...$ and the functions $f(u),\; h(u),\; e(u)$
below are arbitrary, if there is no another notation)

$\varepsilon=+$
\bea &\check{R}(u)=\mathbb{I}+u (f_0 P'^{11}_{\mathcal{I}++}+g_0
P'^{22}_{\mathcal{I}++} +h_0 P'^{12}_{\mathcal{I}++}+e_0
P'^{21}_{\mathcal{I}++}),\label{r1}&\\& \check{R}(u)=f(u)
P'^{11}_{\mathcal{I}++}+g(u) P'^{22}_{\mathcal{I}++} +h(u)
P'^{12}_{\mathcal{I}++}+e(u)
P'^{21}_{\mathcal{I}++}\label{r2x}.&\eea As we can verify, the
matrix (\ref{r2x}) is not invertible and in the standard scheme of
constructing commuting charges via the transfer matrices it is not
usable. But the particular case of that matrix, namely,
\bea & \check{R}(u)=(g(u)+f_0 h(u)) P'^{11}_{\mathcal{I}++}+g(u)
P'^{22}_{\mathcal{I}++} +h(u) P'^{12}_{\mathcal{I}++}+e_0 h(u)
P'^{21}_{\mathcal{I}++} \label{r2u},& \eea
satisfies to $[\check{R}(u),\check{R}(w)]=0$ and hence, the
transfer matrices (as well as monodromy matrices) with different
spectral parameters  constructed by them are also commuting.

 $\varepsilon=-$

 \bea& \check{R}(u)=f(u) \left[P'^{11}_{\mathcal{I}--}\!+\!
P'^{12}_{\mathcal{I}--}\!-\! P'^{22}_{\mathcal{I}--}\!-\!
P'^{21}_{\mathcal{I}--}\right]\!+g(u)\!\left[P'^{12}_{\mathcal{I}--}+
P'^{21}_{\mathcal{I}--}+f_0(P'^{22}_{\mathcal{I}--}+
P'^{21}_{\mathcal{I}--})\right]\!\label{r3x},&\\&
\check{R}(u)=f(u) \left[P'^{11}_{\mathcal{I}--}-
P'^{21}_{\mathcal{I}--}\right] +g(u) \left[
P'^{12}_{\mathcal{I}--}-
P'^{22}_{\mathcal{I}--}\right],\label{r3x1}&\\& \check{R}(u)=f(u)
\left[P'^{11}_{\mathcal{I}--}+ P'^{12}_{\mathcal{I}--}\right]
+g(u) \left[ P'^{22}_{\mathcal{I}--}+
P'^{21}_{\mathcal{I}--}\right].\label{r3x2}&\eea
In the three equations above (\ref{r3x}-\ref{r3x2}) the functions
are not arbitrary, $\frac{f(u)}{g(u)}=u$ or
$\frac{f(u)}{g(u)}=\mathrm{e}^u $. The solutions with the property
$\check{R}(0)=\mathbb{I}$ are the following

\bea&\check{R}(u)=\mathbb{I}+\frac{2(\mathrm{e}^u-1)}{(1+\mathrm{e}^u)\left(g_0^{1/2}-g_0^{-1/2}\right)^2}
\Big[P'^{11}_{\mathcal{I}--}+g_0  P'^{12}_{\mathcal{I}--}-
P'^{22}_{\mathcal{I}--}-g_0^{-1}
P'^{21}_{\mathcal{I}--}\Big],&\label{frg}\\&
\check{R}(u)=\mathbb{I}+u \left( g_0[P'^{11}_{\mathcal{I}--}+
P'^{12}_{\mathcal{I}--}- P'^{22}_{\mathcal{I}--}-
P'^{21}_{\mathcal{I}--}]+h_0[P'^{11}_{\mathcal{I}--}+(1-e_0)
P'^{12}_{\mathcal{I}--}+e_0
P'^{22}_{\mathcal{I}--}]\right).&\label{fr}
 \eea

 Among the
constant solutions we separate the solution
$$\check{R}=c^{2\;2\;2\;2}=\sum_{i,\;\varepsilon=\pm}P'^{ii}_{\mathcal{I}\varepsilon\varepsilon},$$
note that at general $q$  the Casimir operator $c^{2\;2\;2\;2}$
does not satisfy to the YBE. Two another solutions,
\be\label{r3x3}\check{R}=P'^{11}_{\mathcal{I}--}-P'^{22}_{\mathcal{I}--}+
P'^{12}_{\mathcal{I}--}-P'^{21}_{\mathcal{I}--}\quad  \mbox{and}
\quad \check{R}=\sum_{i}P'^{ii}_{\mathcal{I}++}.\ee
are connected with  the
 solutions
$\check{R}^{3\;3}_{1,2}(u)$ taken in the limit $q\to
 i$ (after the multiplication by
$(1+q^2)^2$, i.e. the singular parts) written in the
representation space $V_2\otimes V_2 \otimes V_2 \otimes V_2$. The
first one is the exact $16\times 16$-dimensional analogue of the
mentioned matrices in the limit $q\to i$, the second one is
obtained just by replacing the $c^{3\;3}$- and
$\mathbb{I}^{3\;3}$-matrices by $c^{2\;2\;2\;2}$ and
$\mathbb{I}^{2\;2\; 2\;2}$ in the $\check{R}^{3\;3}_{1,2}(u)$,
which we can denote by $\check{R}^{2\;2\;2\;2}_{1,2}(u)$ (it is
not a solution at general $q$) and then taking the limit $q\to i$
(previously removing the singularities with multiplying by
$(1+q^2)^2$). There is an obvious connection between two matrices
$P'^{11}_{\mathcal{I}--}-P'^{22}_{\mathcal{I}--}+
P'^{12}_{\mathcal{I}--}-P'^{21}_{\mathcal{I}--}\approx\lim_{q\to
i}\Big((P_3\otimes P_3)\check{R}^{2\;2\;2\;2}_{1,2}(u)(P_3\otimes
P_3)\Big)$.

\paragraph{2.} As another class of the solutions
we  consider the matrices with the projectors
$P^{ij}_{\mathcal{I}\varepsilon\varepsilon}$. 
%
\bea\nn
 \check{R}(u)=a\mathbb{I}+f^+(u)
P^{11}_{\mathcal{I}++}+g^+(u) P^{22}_{\mathcal{I}++} + h^+(u)
P^{12}_{\mathcal{I}++}+e^+(u) P^{21}_{\mathcal{I}++}\\+ f^-(u)
P^{11}_{\mathcal{I}--}+g^-(u) P^{22}_{\mathcal{I}--} + h^-(u)
P^{12}_{\mathcal{I}--}+e^-(u) P^{21}_{\mathcal{I}--}. \label{rcp}
\eea
There are few constant solutions with such $R$-matrices.  Putting
$f^+(u)=g^+(u)=e^+(u)=h^+(u)=0$ in (\ref{rcp}) we find no
solutions (constant or spectral parameter dependent) to the YBE.
In contrast to this, when $f^-(u)=g^-(u)=e^-(u)=h^-(u)=0$, there
are numerous solutions, as presented below (\ref{rch}-\ref{rp12}).
Here we represent the spectral parameter dependent solutions
(corresponding constant ones can be obtained as the limits $u\to
\pm \infty$), for which $\check{R}(0)=\mathbf{I}$
\bea&
 \check{R}(u)=
P^{11}_{\mathcal{I}++}+ \mathrm{e}^{2u} P^{22}_{\mathcal{I}++} +
\mathrm{e}^u( P^{11}_{\mathcal{I}--}+ P^{22}_{\mathcal{I}--}).
\label{rch}&\\
&\check{R}(u)=\mathbb{I}+(\mathrm{e}^u
-1) P^{11}_{\mathcal{I}++},\label{rchx}\;\quad 
\check{R}(u)=\mathbb{I}+(\mathrm{e}^u
-1) P^{22}_{\mathcal{I}++}, &\\
&\check{R}(u)=\mathbb{I}+(\mathrm{e}^u-1) P^{11}_{\mathcal{I}++}
+(\mathrm{e}^{-u} -1) P^{22}_{\mathcal{I}++}+f_0
(\mathrm{e}^u-\mathrm{e}^{-u})P^{12/21}_{\mathcal{I}++}.&\label{rchz}\eea

 We can continue the list of  such solutions
presenting  a general solution with $a=1$ and ($f_0,\; g_0$ are
arbitrary)
\bea\label{rgf}
\{f^+(u),\;g^+(u),\;e^+(u),\;h^+(u)\}=\frac{(\mathrm{e}^u-1)}{2\bar{f}_0}\{\pm
g_0+\bar{f}_0,\; \mp g_0+\bar{f}_0,\;\mp 2 f_0,\;\mp
2\},\\\nn\quad \bar{f}_0=\sqrt{4f_0+g_0^2}. \eea
The solutions (\ref{rchx}) as well as solutions like as (below
"$/$" means that all four possibilities are admissible)
\bea
\check{R}(u)=\mathbb{I}+(\mathrm{e}^u -1)
P^{11/22}_{\mathcal{I}++} +e_0
(\mathrm{e}^u-1)P^{12/21}_{\mathcal{I}++}
\eea
 are the particular
cases of the solution (\ref{rgf}).

 Besides of the listed
solutions, there are simple rational solutions also
\bea\check{R}(u)=\mathbb{I}+u\; P^{12/21}_{\mathcal{I}++}
.\label{rp12}\eea

At the end of this subsection, we would like to mention, that our
attempts to find the solutions with the matrices $
 \check{R}(u)=\mathbb{I}+ f^\varepsilon(u)
P^{11}_{\mathcal{I}\varepsilon\varepsilon}+g^\varepsilon(u)
P^{22}_{\mathcal{I}\varepsilon\varepsilon} + h^\varepsilon(u)
P'^{11}_{\mathcal{I}\varepsilon\varepsilon}+e^\varepsilon(u)
P'^{22}_{\mathcal{I}\varepsilon\varepsilon}$, $\varepsilon=\pm$,
where $h^+(u)\neq 0$ or $e^+(u)\neq 0$ for $\varepsilon=+$, bring
us to the conclusion that there is no any solution to the YBE with
such expansion.

\paragraph{3.} Next we observe the solutions with
the projectors $P^{ij}_{\mathcal{I}\varepsilon\bar{\varepsilon}}$. 
Let $\check{R}(0)=\mathbb{I}$.

Here we obtain the following rational solutions
\bea \nn&\check{R}(u)=\mathbb{I}+ u\left(f_0
P^{11}_{\mathcal{I}-+}+g_0 P^{21}_{\mathcal{I}-+} + e_0
P^{21}_{\mathcal{I}+-}+h_0 P^{22}_{\mathcal{I}+-}\right),&\\\nn&
\check{R}(u)=\mathbb{I}+ u\left(f_0 (P^{11}_{\mathcal{I}+-}+
P^{12}_{\mathcal{I}+-}) + e_0
(P^{21}_{\mathcal{I}+-}+P^{22}_{\mathcal{I}+-})+g_0
(P^{11}_{\mathcal{I}-+}-P^{21}_{\mathcal{I}-+})+h_0(
P^{22}_{\mathcal{I}-+}-P^{12}_{\mathcal{I}-+}
)\right),&\\\label{p1}& \check{R}(u)=\mathbb{I}+ u\left(f_0
(P^{11}_{\mathcal{I}+-}+ P^{12}_{\mathcal{I}+-}) + e_0
(P^{21}_{\mathcal{I}+-}+P^{22}_{\mathcal{I}+-})+g_0
P^{11}_{\mathcal{I}-+}+h_0 P^{21}_{\mathcal{I}-+}\right),&\\\nn&
\check{R}(u)=\mathbb{I}+ u\left( f_0 P^{21}_{\mathcal{I}+-}+e_0
P^{22}_{\mathcal{I}+-}+g_0
(P^{11}_{\mathcal{I}-+}-P^{21}_{\mathcal{I}-+})+h_0(
P^{22}_{\mathcal{I}-+}-P^{12}_{\mathcal{I}-+} )\right),&\\\nn&
\check{R}(u)=\mathbb{I}+ u\left(f_0 (2i P^{11}_{\mathcal{I}+-}+ 2i
P^{12}_{\mathcal{I}+-}+P^{12}_{\mathcal{I}-+}-P^{22}_{\mathcal{I}-+})
+\right.&\\\nn&(e_0+2i h_0+2i g_0) P^{21}_{\mathcal{I}+-}+e_0
P^{22}_{\mathcal{I}+-}+g_0 P^{11}_{\mathcal{I}-+}+h_0
P^{21}_{\mathcal{I}-+}\left.\right)\eea
and trigonometric solutions
\bea \nn&\check{R}(u)=\mathbb{I}+
\frac{1-\mathrm{e}^u}{1+\mathrm{e}^u}\left(\pm 2
P^{12}_{\mathcal{I}+-}\mp i P^{12}_{\mathcal{I}-+} +
f_0(P^{11}_{\mathcal{I}-+}-2 i
P^{22}_{\mathcal{I}+-})+g_0(P^{21}_{\mathcal{I}-+}+2i
P^{21}_{\mathcal{I}+-})\right.&\\&\left.+e_0(
P^{22}_{\mathcal{I}-+}-2i P^{11}_{\mathcal{I}+-}-2i
P^{12}_{\mathcal{I}+-}-P^{12}_{\mathcal{I}-+})\right).
&\label{pk}\eea
 Some solutions in (\ref{p1}) can coincide
one with other for the particular choices of the arbitrary
parameters $f_0,\; g_0,\;e_0$ and $h_0$.

The solutions with the projectors
${P'}^{ij}_{\mathcal{I}\varepsilon\bar{\varepsilon}}$ are quite
similar to (\ref{p1}, \ref{pk}).
\bea &\check{R}(u)=\mathbb{I}+ u\left(f_0
P'^{11}_{\mathcal{I}+-}+g_0 P'^{12}_{\mathcal{I}+-}+e_0
P'^{12}_{\mathcal{I}-+}+h_0 P'^{22}_{\mathcal{I}-+}\right),&\nn\\
&\check{R}(u)=\mathbb{I}+ u\left(f_0
(P'^{11}_{\mathcal{I}+-}+P'^{12}_{\mathcal{I}+-})+g_0(
P'^{21}_{\mathcal{I}+-}+P'^{22}_{\mathcal{I}+-}) + e_0(
P'^{21}_{\mathcal{I}-+}-P'^{11}_{\mathcal{I}-+})+h_0(
P'^{22}_{\mathcal{I}-+}-P'^{12}_{\mathcal{I}-+})\right),&\nn\\&
\check{R}(u)=\mathbb{I}+ u\left(f_0 (P'^{11}_{\mathcal{I}+-}+
P'^{12}_{\mathcal{I}+-}) + e_0
(P'^{21}_{\mathcal{I}+-}+P'^{22}_{\mathcal{I}+-})+g_0
P^{12}_{\mathcal{I}-+}+h_0
P^{22}_{\mathcal{I}-+}\right),&\label{p'}\\\nn&
\check{R}(u)=\mathbb{I}+ u\left( f_0 P'^{12}_{\mathcal{I}+-}+e_0
P'^{11}_{\mathcal{I}+-}+g_0
(P'^{11}_{\mathcal{I}-+}-P'^{21}_{\mathcal{I}-+})+h_0(
P'^{22}_{\mathcal{I}-+}-P'^{12}_{\mathcal{I}-+} )\right),&\\\nn&
\check{R}(u)=\mathbb{I}+ u\left(f_0 (P'^{11}_{\mathcal{I}-+}-
P'^{21}_{\mathcal{I}-+}+2i P'^{21}_{\mathcal{I}+-}+2i
P'^{22}_{\mathcal{I}+-}) +\right.&\\\nn& (e_0+2i h_0+2i g_0)
P'^{11}_{\mathcal{I}+-}+e_0 P'^{12}_{\mathcal{I}+-}+g_0
P'^{22}_{\mathcal{I}-+}+h_0
P'^{12}_{\mathcal{I}-+}\left.\right)&\\\nn&
\check{R}(u)=\mathbb{I}+
\frac{1-\mathrm{e}^u}{1+\mathrm{e}^u}\left(\pm i
P'^{21}_{\mathcal{I}-+}\pm 2 P'^{21}_{\mathcal{I}+-} + f_0(2i
P'^{11}_{\mathcal{I}+-}+
P'^{22}_{\mathcal{I}-+})+g_0(P'^{12}_{\mathcal{I}-+}-2i
P'^{12}_{\mathcal{I}+-})+\right.&\\&\left.e_0(2i
P'^{22}_{\mathcal{I}+-}+ P'^{11}_{\mathcal{I}-+}+2i
P^{21}_{\mathcal{I}+-}-P'^{21}_{\mathcal{I}-+})\right).&
\label{p3}\eea

Of course, consideration of the other possible structures of the
$R$-matrices with different combinations of the projector
operators also will give new solutions.

\paragraph{Note.} Here we do not display  all the solutions $R^{44}(u)$ to
the YBE at general $q$ or at roots of unity. The full list of the
solutions are obtained for some definite cases (grouped in the
marked paragraphs 1-3, for the last two cases provided
$\check{R}(0)=\mathbb{I}$). However the presented results at roots
of unity demonstrate the existence of the solutions which cannot
be obtained from the solutions at general $q$. The plain evidence
of it is the presence in the solutions of the projectors
($P'^{ij}_{\mathcal{I}++},\;
P'^{ij}_{\mathcal{I}\varepsilon\bar{\varepsilon}}$), which (wholly
or separately) do not coincide with the limit $q\to i$ of any
linear combination of the projectors existing at general $q$. The
peculiarities of the obtained solutions, i.e. their large number
and variety
 (constant ones, solutions with rational, exponential or
trigonometric dependence on the spectral parameter, solutions
containing arbitrary functions), existence of the rich number of
 arbitrary parameters, argue the novelty of their nature.

\section{Chain models corresponding to the solutions.}

\addtocounter{section}{0}\setcounter{equation}{0}

This section is devoted to the study of the integrable models
which can be defined using the YBE solutions described above, via
the transfer matrix approach \cite{ybe,rbaxter,grs}.

 Let us define quantum
space of a chain with $N$ sites as $\mathbf{A}_N=A_1\otimes A_2
\cdots \otimes A_N$, where $A_i$ is the vector space corresponding
to the $i$-th site, and serves as a representation space of the
algebra $sl_q(2)$. If to construct transfer matrix $\tau(u)={tr}_a
\prod_i R_{ai}(u)$, with the operators $R_{ai}(u)$ which
 act on the vector spaces $A_a\otimes
A_i$, and coincide with the solutions to the YBE obtained at roots
of unity, then the resulting quantum chain model with the
Hamiltonian operator defined as the first logarithmic derivative
of the
transfer matrix at the normalization point $u_0$ ($\check{R}(u_0)=\mathbb{I}$) 
can be treated as an extended XXZ model at
roots of unity. We intend to investigate the case when $q=i$, i.e.
the case of the extended XX models.

We take
$A_i=\left[\mathcal{I}^{(4)}_{\{3,1\}}\right]_i=[V_2]_{2i}\otimes
[V_2]_{2i+1}$. The solution given by the expression (\ref{rr4})
corresponds to the ordinary XX model, with the following lattice
Hamiltonian ($k\equiv 2i-1$)
\bea \nn&H_{XX}=J\sum_{k,\;\Delta k=2}^{2N}
\Big(\sigma^+_{k}\sigma^-_{k+1}+\sigma^-
_{k}\sigma^+_{k+1}+2(\sigma^+_{k+1}\sigma^-_{k+2}+\sigma^-
_{k+1}\sigma^+_{k+2})&\\\nn&+\sigma^+_{k+3}\sigma^-_{k+4}+\sigma^-
_{k+3}\sigma^+_{k+4}+\frac{i}{2}(\sigma^z_k+\sigma^z_{k+1}-\sigma^z_{k+3}-\sigma^z_{k+4}
)\Big)&\\&= J\sum_{k,\;\Delta k=1}^{2N}
\Big(\sigma^+_k\sigma^-_{k+1}+\sigma^-
_{k}\sigma^+_{k+1}+\frac{i}{2}(\sigma^z_{k}-\sigma^z_{k+1}
)\Big).& \label{hxx}\eea
Here the cyclic boundary conditions $\sigma^k_1=\sigma^k_{2N+1}$
and $\sigma^k_2=\sigma^k_{2N+2}$ (with
$\sigma^+=\left(\ba{cc}0&1\\0&0\ea\right),\;\sigma^-=\left(\ba{cc}0&0\\1&0\ea\right),
\;\sigma^z=\left(\ba{cc}1&0\\0&-1\ea\right)$) are imposed, and the
terms with $\sigma^z_i$-operators, ensuring $sl_i(2)$ symmetry,
are disappeared in the entire expression. The same Hamiltonian can
be obtained, as it is well known, from the fundamental $R^{2\;
2}(u)$-matrix at $q=i$. The appearing of the coupling constant $J$
in (\ref{hxx}) mathematically reflects the freedom of the scaling
of the spectral parameter $u$. It must be real, in order to keep
the hermicity of the Hamiltonian operator. But for the cases
brought below, when the hermicity is broken, there is no general
condition on $J$.

\subsection{Extended XX models: non-Hermitian Hamiltonian operators.}

Now let us write the Hamiltonian operators corresponding to the
new obtained solutions. We shall observe few of them, so that to
touch on all the obtained types of the  solutions. We shall start
with the construction of the model given by the $R$-matrix
 (\ref{rcf}). The simplest case, which corresponds to
the sum of the unity and Casimir operators, gives the following
expression
\bea \nn&H^c= \sum_{k,\;\Delta k=2}^{2N} \Big(\sigma^+_k
\sigma^-_{k+3}+\sigma^-_{k}\sigma^+_{k+3}+i
\sigma^z_k(\sigma^+_{k+1}
\sigma^-_{k+3}+\sigma^-_{k+1}\sigma^+_{k+3})-i (\sigma^+_k
\sigma^-_{k+2}+\sigma^-_{k}\sigma^+_{k+2})\sigma^z_{k+3}&\\\nn&
+\sigma^z_k(\sigma^+_{k+1}
\sigma^-_{k+2}+\sigma^-_{k+1}\sigma^+_{k+2})\sigma^z_{k+3}-(\sigma^+_k
\sigma^-_{k+1}+\sigma^-_{k}\sigma^+_{k+1})
\sigma^z_{k+2}\sigma^z_{k+3}-\sigma^z_k
\sigma^z_{k+1}(\sigma^+_{k+2}
\sigma^-_{k+3}+\sigma^-_{k+2}\sigma^+_{k+3})&\\&
+\frac{i}{2}(\sigma^z_k \sigma^z_{k+1}\sigma^z_{k+3}+
\sigma^z_{k+1}\sigma^z_{k+2} \sigma^z_{k+3}-\sigma^z_k
\sigma^z_{k+1}\sigma^z_{k+2} -\sigma^z_{k}\sigma^z_{k+2}
\sigma^z_{k+3})\Big).&\label{hc1} \eea
 And apparently,
the Hamiltonian (\ref{hc1}) in the representation of the scalar
fermions, evaluated by means of the Jordan-Wigner transformations,
\bea \sigma^{+}_i=c_i\prod_{j=1}^{i-1} (1-2 c^+_j c_j),\quad
\sigma^{-}_i=c^{+}_i\prod_{j=1}^{i-1} (1-2 c^+_j c_j),\quad
\sigma^{z}_i= 1-2 c^+_i c_i,\eea
see as example \cite{grs,shs}, contains interaction terms up to
the sixth power of the fermion operators and, hence, is not
free-fermionic as it was in the case (\ref{hxx}). Also, it
contains non-Hermitian terms. Note, that the next to nearest
Hamiltonian derived from the fundamental $R^{2\; 2}(u)$-matrix
(i.e. second logarithmic derivative of the transfer matrix)
contain terms like $\sigma^{\pm}_i\sigma^z_{i+1}\sigma^{\mp}_{i+2}
$ ($=c^+_{i}c_{i+2}$ or $c^+_{i+2}c_{i}$), i.e. describes free
fermions.

It is interesting to present the Hamiltonian operators
corresponding to the new solutions (with the $R$-matrices which
cannot be obtained as the limits at roots of unity of the matrices
at general $q$). Such matrices are,  as example,
$\check{R}^{12/21}(u)=\mathcal{I}+u P^{12/21}_{++}$ (\ref{rp12}).
Hamiltonian operators corresponding to them are (in the spin and
fermionic representations)
\bea H^{12}_{++}= J\sum_{k,\;\Delta k=2}^{2N}\Big(\sigma^+_{k+1}
\sigma^+_{k+2}-i\sigma^+_{k}\sigma^z_{k+1}\sigma^+_{k+2}-
\sigma^+_{k} \sigma^+_{k+1}\Big)=\label{h12}\\\nn
J\sum_{i}^{N}\Big(\sigma^+_{2i}
\sigma^+_{2i+1}-i\sigma^+_{2i-1}\sigma^z_{2i}\sigma^+_{2i+1}-
\sigma^+_{2i-1} \sigma^+_{2i} \Big)
 \Rightarrow J\sum_{i}^{N}\Big(
c_{2i+1}c_{2i}-ic_{2i+1}c_{2i-1}- c_{2i}c_{2i-1} \Big),\\
H^{21}_{++}= J\sum_{k,\;\Delta k=2}^{2N}\Big(\sigma^{-}_{k+1}
\sigma^-_{k+2}-i\sigma^-_{k}\sigma^z_{k+1}\sigma^-_{k+2}-
\sigma^-_{k} \sigma^-_{k+1}\Big)=\label{h21}\\\nn
J\sum_{i}^{N}\Big(\sigma^{-}_{2i}
\sigma^-_{2i+1}-i\sigma^-_{2i-1}\sigma^z_{2i}\sigma^-_{2i+1}-
\sigma^-_{2i-1} \sigma^-_{2i}\Big)\Rightarrow
J\sum_{i}^{N}\Big(c^{+}_{2i}
c^+_{2i+1}-ic^+_{2i-1}c^+_{2i+1}-c^+_{2i-1}c^+_{2i}\Big).\eea
As we see they both are non-Hermitian free-fermionic operators.

Another Hamiltonian operators resulted  from the new solutions,
can be found from the matrices (\ref{rcf}, \ref{r1}, \ref{frg},
\ref{fr}, \ref{rcp}-\ref{pk}).

Among the mentioned solutions we can see that the matrix
(\ref{rchz}) at small $u$ and at $f_0=0$ takes the form
$\check{R}(u)=\mathbb{I}+u(P^{11}_{++}-P^{22}_{++})$, and hence
the corresponding Hamiltonian writes as
 \bea H_{++}= J\sum_{k,\;\Delta k=2}^{2N}\Big(i(\sigma^+_{k}
\sigma^-_{k+1}+\sigma^-_{k} \sigma^+_{k+1}-\sigma^+_{k+1}
\sigma^-_{k+2}-\sigma^-_{k+1}
\sigma^+_{k+2})-\sigma^+_{k}\sigma^z_{k+1}\sigma^-_{k+2}-\sigma^-_{k}\sigma^z_{k+1}\sigma^+_{k+2}+
\sigma^z_{k+1}\Big). \eea
The corresponding fermionic representation of the Hamiltonian
looks like as follows
\bea H^{f}_{++}\!=\!J\sum_{i}^{N}\! \Big(i(c^{+}_{2i-1}
c_{2i}+c^{+}_{2i}c_{2i-1}-c^{+}_{2i} c_{2i+1}-c^{+}_{2i+1}c_{2i}
)-c^{+}_{2i-1}c_{2i+1}-c^{+}_{2i+1}c_{2i-1}+
1-2c^{+}_{2i}c_{2i}\Big).\label{hcc} \eea
If in (\ref{rchz}) $f_0\neq 0$, then the additional term for the
case of $P^{12}_{\mathcal{I}++}$ writes as $2 f_0
J\sum_{i}^{N}(\sigma^+_{2i} \sigma^+_{2i+1}-\sigma^+_{2i-1}
\sigma^+_{2i}-i \sigma^+_{2i-1}\sigma^z_{2i} \sigma^+_{2i+1})$ or,
in the fermionic representation, $2 f_0
J\sum_{i}^{N}(c_{2i+1}c_{2i}+c_{2i-1} c_{2i}+i c_{2i-1}
c_{2i+1})$. For obtaining the case of $P^{21}_{\mathcal{I}++}$ the
operators $\sigma^+_{i}$ and $c_{i}$ one must change by the
operators $\sigma^-_{i}$ and $c^+_i$.

 In the graphical
representation the Hamiltonian operators (\ref{h12}, \ref{h21},
\ref{hcc}) can be depicted more apparently on the lattices, where
the odd and even numbered spins are shown on two different chains.
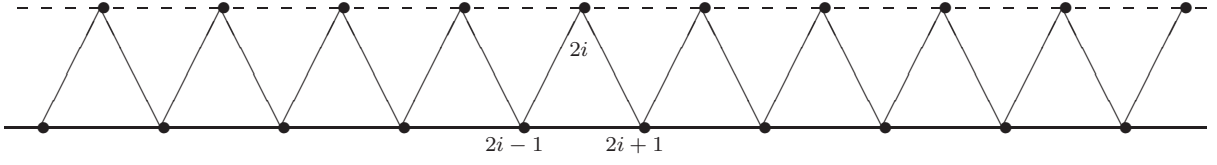
\begin{figure}[t]
\vspace{0.5cm}
\begin{picture}(100,25)(0,0)
\unitlength=9pt
 \multiput(0,0)(0,5){2}{\line(1,0){50}}
 {\color{white}{\multiput(0,5)(1,0){50}{\line(1,0){0.5}}}}
 \put(20,-1){\scriptsize$2i-1$}%
 \put(25,-1){\scriptsize$2i+1$}%
 \put(23.5,3){\scriptsize$2i$}%
  \multiput(1.5,0)(5,0){10}{\line(1,2){2.5}}
   \multiput(6.5,0)(5,0){9}{\line(-1,2){2.5}}
   \multiput(1,-0.3)(2.5,5){2}{\multiput(0,0)(5,0){10}{$\;\bullet$}}
\end{picture}
 \caption{Graphical representations  of the spin-chain Hamiltonians (\ref{h12}, \ref{h21}, \ref{hcc}).}\label{fr1}
\end{figure}
In  Fig. \ref{fr1} the spin (or fermionic) variables are attached
on the sites noted by the dots. The next-to-nearest Hamiltonians
(\ref{h12}, \ref{h21}, \ref{hcc}) contain hopping terms only along
the thick lines of the figure.

The particular solutions of (\ref{frg}) and (\ref{fr}),
$$\check{R}^{\pm}(u)=\mathbb{I}+u
\left(P'^{11}_{\mathcal{I}--}- P'^{22}_{\mathcal{I}--}\pm(
P'^{12}_{\mathcal{I}--}- P'^{21}_{\mathcal{I}--})\right),$$
 give rise to  "factorized" Hamiltonian operators, 
which look like as
 \bea\label{hf}
H^{factor+}_{--}=\sum_{k,\;\Delta k=2}^{2N}
h_{k,k+1}h_{k+2,k+3}=\\\nn J^+\sum_{k,\;\Delta k=2}^{2N}
\Big(\sigma^{+}_{k}
\sigma^-_{k+1}+\sigma^+_{k+1}\sigma^-_{k}+\frac{i}{2}(\sigma^z_{k}-\sigma^z_{k+1})\Big)\Big(\sigma^{+}_{k+2}
\sigma^-_{k+3}+\sigma^+_{k+3}\sigma^-_{k+2}+\frac{i}{2}(\sigma^z_{k+2}-\sigma^z_{k+3})\Big),\\\label{hff}
H^{factor-}_{--}=\sum_{k,\;\Delta k=2}^{2N}h_{k,k+3}
h_{k+1,k+2}=\\\nn J^-\sum_{k,\;\Delta k=2}^{2N}\Big(\sigma^{+}_{k}
\sigma^-_{k+3}+\sigma^+_{k+3}\sigma^-_{k}+\frac{i}{2}(\sigma^z_{k}-\sigma^z_{k+3})\Big)\Big(\sigma^{+}_{k+1}
\sigma^-_{k+2}+\sigma^+_{k+2}\sigma^-_{k+1}+\frac{i}{2}(\sigma^z_{k+1}-\sigma^z_{k+2})\Big).
\eea
Note, that the Hamiltonian  of the ordinary XX model is
$\sum_{i}^{2N} h_{i,i+1}$ and the second Hamiltonian (second
logarithmic derivative of the transfer matrix) is proportional to
$\sum_{i}^{2N} [h_{i,i+1}, h_{i+1,i+2}]$ \cite{grs}. In the
fermionic representation both of them contain only quadratic terms
(describe free fermions), in the contrast of the Hamiltonian
operators (\ref{hf}) and (\ref{hff}), which describe fermions with
quartic interaction terms. Note also, that the term
$h_{i,j}=\sigma^{+}_{i}
\sigma^-_{j}+\sigma^+_{j}\sigma^-_{i}+\frac{i}{2}(\sigma^z_{i}-\sigma^z_{j})$
is simply the Casimir operator $c^{2\;2}$ defined on
$[V_2]_{i}\otimes [V_2]_{j}$. And, particularly, the operator
(\ref{hf}) can be represented also as
$H^{factor+}_{--}=\sum_{i}^{N}
h_{2i,2i+1}h_{2i+2,2i+3}=\sum_{i}^{N} [c^{2\;2}]_i
[c^{2\;2}]_{i+1}$, being interpreted as a quadratic interaction
between two nearest-neighbored four-dimensional indecomposable
vector spaces.

\begin{figure}[t]
\vspace{0.5cm}
\begin{picture}(100,25)(0,0)
\unitlength=9pt
 \multiput(0,0)(0,5){2}{\line(1,0){50}}
 {\color{white}{\multiput(0,5)(1,0){50}{\line(1,0){0.5}}}}
  {\color{white}{\multiput(0,0)(1,0){50}{\line(1,0){0.5}}}}
 \put(20,-1){\scriptsize$2i-1$}%
 \put(25,-1){\scriptsize$2i+1$}%
 \put(23.7,5.5){\scriptsize$2i$}%
 \put(28,5.5){\scriptsize$2i\!+\!2$}%
  \multiput(1.5,0)(5,0){10}{\line(1,2){2.5}}
    \multiput(5,2.3)(5,0){9}{$\mathbf{\otimes}$}
   \multiput(1,-0.3)(2.5,5){2}{\multiput(0,0)(5,0){10}{$\;\bullet$}}
\end{picture}
 \caption{Graphical representations  of the spin-chain Hamiltonian (\ref{hf}).}\label{fr2}
\end{figure}
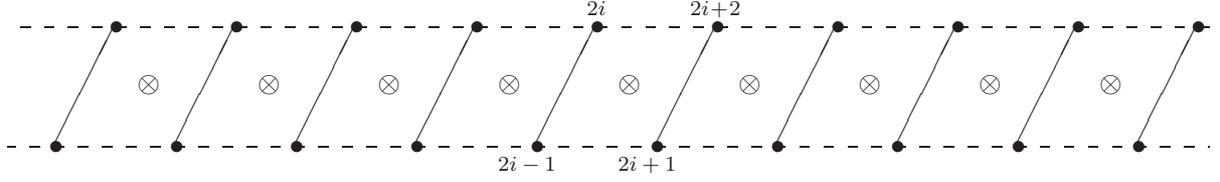
\begin{figure}[t]
\vspace{0.5cm}
\begin{picture}(100,32)(0,0)
\unitlength=9pt
 \multiput(0,0)(0,5){2}{\line(1,0){50}}
 {\color{white}{\multiput(0,5)(1,0){50}{\line(1,0){0.5}}}}
  {\color{white}{\multiput(0,0)(1,0){50}{\line(1,0){0.5}}}}
 \put(20,-1){\scriptsize$2i-1$}%
 \put(25,-1){\scriptsize$2i+1$}%
 \put(23.7,5.5){\scriptsize$2i$}%
 \put(28,5.5){\scriptsize$2i\!+\!2$}
   \multiput(6.5,0)(5,0){9}{\line(-1,2){2.5}}
    \multiput(1.5,0)(5,0){9}{\line(3,2){7.5}}
    \multiput(5,2.3)(5,0){9}{$\mathbf{\otimes}$}
   \multiput(1,-0.3)(2.5,5){2}{\multiput(0,0)(5,0){10}{$\;\bullet$}}
\end{picture}
 \caption{Graphical representations  of the spin-chain Hamiltonian (\ref{hff}).}\label{fr3}
\end{figure}
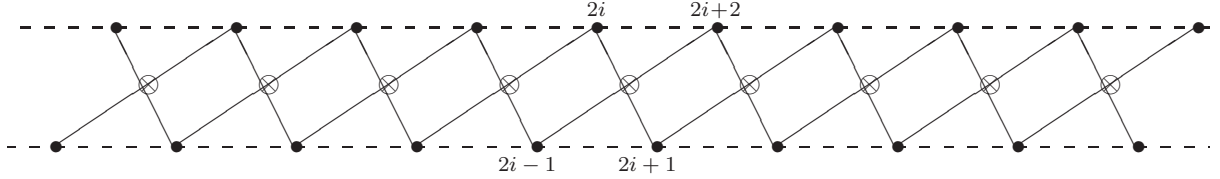
In  Figs. \ref{fr2}, \ref{fr3} we represent the quartic
Hamiltonians (\ref{hf}) and (\ref{hff}) in a graphical way: the
local interactions take place between the spins (fermions)
disposed on the four neighbored sites around the marked centers,
with interaction terms presented by the products of two hopping
terms $h_{ij}$ along two thick lines, which are in the close
vicinity of each center(Fig. \ref{fr2}) or are crossed in the
centers (Fig. \ref{fr3}).

 For completeness let us give also some Hamiltonian operators
corresponding to  the solutions (\ref{p1}-\ref{p3}). The second
solution of (\ref{p1}) with the choice of the parameters
$\{f_0,e_0,g_0,h_0\}=J_0\{1,1,i/2,i/2\}$ leads to the following
Hamiltonian
\bea H_{+-}= J\sum_{k,\;\Delta k=2}^{2N}\Big(\sigma^+_{k}
\sigma^-_{k+1}+\sigma^-_{k}
\sigma^+_{k+1}+\frac{i}{2}(\sigma^z_{k}-\sigma^z_{k+1})-\\\nn
(\sigma^+_{k+1} +i\sigma^-_{k+1}+(\sigma^-_{k}-i
\sigma^+_{k})\sigma^z_{k+1})(\sigma^-_{k+2}+i
\sigma^z_{k+2}\sigma^-_{k+3})\Big). \eea
In the fermionic representation it is a non-Hermitian free
fermionic operator
\bea H^{f}_{+-}\!= J\!\!\!\sum_{i,\;\Delta i=2}^{2N}\!\Big(c^+_{k}
c_{k+1}+ c^+_{k+1}c_{k}+i(c^+_{k+1} c_{k+1}-c^+_{k} c_{k})-
(c^{+}_{k}+i c^{+}_{k+1}-c_{k+1}+i c_{k})(c^{+}_{k+2}+i
c^{+}_{k+3})\Big).\label{hxy}\eea
This Hamiltonian by its structure (as well as the operators
(\ref{h12}) and (\ref{h21})) resembles rather the Hamiltonian of
the XY model.

A similar Hamiltonian operator we can found from the solutions
(\ref{p'}), taking in the second matrix the following parameters
$\{f_0,e_0,g_0,h_0\}=J'_0\{1,1,i/2,-i/2\}$,
\bea {H'}_{+-}= J\sum_{k,\;\Delta k=2}^{2N} \Big(\sigma^+_{k}
\sigma^-_{k+1}+\sigma^-_{k}
\sigma^+_{k+1}+\frac{i}{2}(\sigma^z_{k}-\sigma^z_{k+1})-\\\nn
(\sigma^-_{k+1} -i\sigma^+_{k+1}-(\sigma^+_{k}+i
\sigma^-_{k})\sigma^z_{k+1})(\sigma^+_{k+2}+i
\sigma^z_{k+2}\sigma^+_{k+3})\Big). \eea
The corresponding  fermionic representation is
\bea {H'}^{f}_{\!\!\!+-}\!= J\!\!\!\sum_{k,\;\Delta
k=2}^{2N}\!\!\!\Big(c^+_{k} c_{k+1}+ c^+_{k+1}c_{k}+i(c^+_{k+1}
c_{k+1}-c^+_{k} c_{k})\!-\! (c^{+}_{k+1}-i c^{+}_{k}+c_{k}+i
c_{k+1})(c_{k+2}+i c_{k+3})\!\Big).\label{hyx} \eea
 In the last examples given above we have dealt with the
 Hamiltonian functions which are homogeneous polynomials in
 respect of the fermionic operators (homogeneous polynomials of degree two (\ref{h12}, \ref{h21}, \ref{hcc}, \ref{hxy},
 \ref{hyx})- containing only kinetic terms,
 or of degree four (\ref{hf}, \ref{hff})- only interaction
 terms).
 It is conditioned by our aim
 to choose more symmetric matrices among the YBE solutions. But of
 course, a large number of the solutions correspond to
 non-homogeneous Hamiltonians. The fermionic representation of the $H$ in (\ref{hc1})
 contains terms with the
 second, fourth and sixth powers of the operators.
  As an illustration of the Hamiltonian with the 
 four-fermionic interaction term together with a
 kinetic term, 
 we can point the following
 Hamiltonian operators, corresponding to the simple solutions $\check{R}(u)=\mathbb{I}+u
 P^{11}_{\mathcal{I}-+}$, $\check{R}(u)=\mathbb{I}+u
 P^{21}_{\mathcal{I}-+}$ or $\check{R}(u)=\mathbb{I}+u
 \left(P^{11}_{\mathcal{I}-+}-P^{21}_{\mathcal{I}-+}+i(\Delta-2)\left[ P^{21}_{\mathcal{I}+-}+P^{22}_{\mathcal{I}+-}\right]\right)$
 (see
 (\ref{p'})). For the last one the corresponding
 fermionic Hamiltonian is the following
\bea\nn& H^{f}_{+-,\Delta}=J\sum_{i=1}^N\Big(-2(c_{2i-1}+i
c_{2i})(c_{2i+1}+i c_{2i+2})+&\\&
\Delta\left[h_{2i-1,2i}c_{2i+1}c_{2i+2}+(i c_{2i-1}^+c_{2i-1}
c_{2i}+c_{2i-1}c^{+}_{2i}c_{2i})(c_{2i+1}+ic_{2i+2})\right]
\Big).& \eea

\paragraph{Note.}
Taking into account that the local terms of the obtained new
Hamiltonians connect two pairs of the neighboring
spin-$\frac{1}{2}$  states (sometimes they restrict to three-spin
interactions, as in (\ref{h12}, \ref{h21}, \ref{hcc})), reflecting
the composite structure of the states on which the $R$-matrices
are defined, one could relate the obtained models to those ones,
being highly exploited in the strongly correlated systems, such as
the dimer models, ladder (or zigzag) models. A general
disadvantage which inheres in the most of the discussed
Hamiltonian operators is their non-hermicity. The quadratic in
terms of the fermionic operators (i.e. free fermionic) Hamiltonian
operators describe integrable models a priori, as the Fourier
transformation allows to define the full eigen-system of such
models. Hence, the Hermitian parts ($\frac{1}{2}[H+H^{+}],\;
\frac{1}{2i}[H-H^{+}]$) of a quadratic Hamiltonian also describe
integrable models. But now they are fully diagonalizable and have
real spectra, being in general with no $sl_i(2)$ symmetry (the
Hamiltonian operator $H^+$ acquires the symmetry of the algebra
$sl_{-i}(2)$, so the resulting Hamiltonian operators
$\frac{1}{2}[H+H^{+}],\; \frac{1}{2i}[H-H^{+}]$ are the
combinations of the invariant operators in respect of $sl_i(2)$
and $sl_{-i}(2)$). As concerns the Hamiltonian operators with
quartic and higher interactions, in each particular case there is
need to check the integrability of the models defined by the
Hermitian parts of the Hamiltonians.

And at the end of this section we would like to touch on the
spectra of the discussed models with the free-fermionic behaviour.
To obtain physically justified results and in order to deal with
permissible transformations of the fermionic variables, we
consider the Hermitian parts of the Hamiltonian operators.
Particularly, for the fermionic $H$  described
 in (\ref{hcc}), in
the Fourier basis of the chain discrete momenta,
\bea c_{2i}=\frac{1}{\sqrt{N}}\sum_{p=1}^{2N}e^{-\imath\frac{ \pi
(2i) p}{N}}c_{1p},\;\;\;\quad \qquad
c_{2i+1}=\frac{1}{\sqrt{N}}\sum_{p=1}^{2N}e^{-\imath\frac{ \pi
(2i+1) p}{N}}c_{2p},\eea
 the models with the Hamiltonian operators $\frac{1}{2}[H+H^{+}]$ and $\frac{1}{2i}[H-H^{+}]$,
acquire the following energy spectra, correspondingly,
$\{1,\;2\cos{[\frac{2\pi p}{N}]}\}$ and $\{ \pm
\sin{[\pi\frac{p}{N}]}\}$,
 $0\leq
p< N$. The Hermitian parts of the Hamiltonian operators
(\ref{h12}), (\ref{h21}) have the eigenvalues, symmetric in
respect of the origin. They are
$\{\pm\cos{[\pi\frac{p}{N}]}\left(\sin{[\pi\frac{p}{N}]}\pm
\sqrt{1+\sin{[\pi\frac{p}{N}]}^2}\right)\}$ and
$\{\pm\cos{[\pm\pi\frac{p}{N}]}\}$ respectively,  and  here the
eigenvectors are the combinations of the states with opposite
momenta, $c_{1p},\; c_{2p},\;c^+_{1(N-p)},\; c^+_{2(N-p)}$, $0\leq
p< N/2$ \cite{shs}.

\section{Treating
 of the indecomposable representations in the context of the
dynamics of the systems. Non-unitary evolution operators.}
\addtocounter{section}{0}\setcounter{equation}{0}

In this section we want to observe the models with $sl_q(2)$ (as
well as $osp(1|2)_q$) symmetry at roots of unity from another
aspect. As we have seen the Hamiltonian operators which are
constructed taking into account 
the indecomposable
states are
non-Hermitian. It means that 
the evolution matrices of the corresponding models appear to be
non-unitary. But in the recent decades there are numerous
investigations of the systems with non-Hermitian Hamiltonians
\cite{Bender} and there is a chance that investigation of the new
integrable models at roots of unity is not a pure mathematical
analysis only.

The specific, peculiar character of the Hamiltonian operators at
roots of unity consists of the presence of the indecomposable
representations in the spectrum of the eigenstates. Let us observe
the dynamics of such Hamiltonian systems. Suppose we have a chain
with $2N$ sites with Hamiltonian e.g. (\ref{hc1}). Let us consider
the simplest case, when $N=1$. The periodic boundary conditions
imply $\sigma_3=\sigma_1,\;\sigma_4=\sigma_2$. After careful
calculations we are coming to the following two-site Hamiltonian
(with the normalized coefficient $J\to J/4$)
$$ H=J h_{1,2}=J\Big(\sigma^{+}_{1}
\sigma^-_{2}+\sigma^+_{2}\sigma^-_{1}+\frac{i}{2}(\sigma^z_{1}-\sigma^z_{2})\Big).
$$
On the four-dimensional space $V_2\otimes V_2$ this operator has
the matrix form
\bea
H=J\left(\ba{cccc}0&0&0&0\\0&i&1&0\\0&1&-i&0\\0&0&0&0\ea\right).\label{hn1}
\eea
The vectors
$|v_+\rangle={\scriptsize{\left(\ba{c}1\\0\\0\\0\ea\right)\equiv\left(\ba{c}1\\0\ea\right)\otimes\left(\ba{c}1\\0\ea\right)}}$,
${\scriptsize{|v_-\rangle=\left(\ba{c}0\\0\\0\\1\ea\right)\equiv\left(\ba{c}0\\1\ea\right)\otimes\left(\ba{c}0\\1\ea\right)}}$
 and $|v_0\rangle=\frac{1}{\sqrt{2}}{\scriptsize{\left(\ba{c}0\\1\\-i\\0\ea\right)}}\\\equiv\frac{1}{\sqrt{2}}{\scriptsize{
 \left(\ba{c}1\\0\ea\right)\otimes\left(\ba{c}0\\1\ea\right)}}-i
 \frac{1}{\sqrt{2}}{\scriptsize{
 \left(\ba{c}0\\1\ea\right)\otimes\left(\ba{c}1\\0\ea\right)}}$ are
 the eigenstates of the Hamiltonian (\ref{hn1}) with the
 eigenvalue $0$. Any state $|u_0\rangle=
\frac{\gamma}{\sqrt{2}}
{\scriptsize{\left(\ba{c}0\\1\ea\right)\otimes\left(\ba{c}1\\0\ea\right)}}+\alpha|v_0\rangle$
with arbitrary $\alpha$ satisfies to the relation $H\cdot
|u_0\rangle=J \gamma|v_0\rangle$. If to choose
$|u_0\rangle=\frac{e^{i\theta}}{\sqrt{2}}{\scriptsize{
\left(\ba{c}0\\1\\i\\0\ea\right)}}$ (with $\theta$ to be a real
number), then the scalar product defined as $(v^+,w)=(\langle
v|)^\ast|w\rangle$ provides for the orthogonal and normalized
vectors: $(v_\varepsilon^+,v_\eta)=\delta_{\varepsilon\eta},
\;(v_\varepsilon^+,u_0)=0,\;(u_0^+,u_0)=1$, where
$\varepsilon,\;\eta=+,-,0$. Note, that the ordinary scalar product
$(v,w)=\langle v||w\rangle$ (here and in the Appendix we have
denoted by $\langle v|$ the transposed vector $(|v\rangle)^\tau$,
without complex conjugation, in contrast to the usual convention,
where $\langle v|$ means Hermitian conjugation) gives $
(v_0,v_0)=0$ (the vector with zero norm in the indecomposable
representation). In the quantum theory the definition $(v^+,w)$ is
used for measuring the probability of the system to occupy the
given state.

Let us observe how the time evolution flows for the mentioned
states. Usually  considering the non-Hermitian models the authors
try to avoid the problems coming with the non-unitary evolution
matrices and the time-dependent norm \cite{Bender,fring}. Let us
see, what we shall have making a straightforward analysis. The
solutions of the Shr$\mathrm{\ddot{o}}$dinger equation with the
Hamiltonian (\ref{hn1}) are the following time-dependent states:
$|v_\varepsilon(t)\rangle=| v_\varepsilon\rangle$,
$|u_0(t)\rangle=|u_0\rangle-i t J \gamma | v_0\rangle$. Note, that
the norm of the state $|u_0(t)\rangle$ changes with time as
follows $(u_0(t)^+,u_0(t))=1+4 |J t|^2$ (we use the vector
$|u_0\rangle$ fixed above). Hence the normalized state
$$|\bar{u}_0(t)\rangle=\frac{|u_0(t)\rangle}{\sqrt{(u_0(t)^+,u_0(t))}}=
\frac{|u_0\rangle+2J e^{i\theta}t | v_0\rangle}{\sqrt{1+4 |J
t|^2}}$$
 in the limit $t\to\infty$ becomes $e^{i\theta}\frac{J}{|J|}|v_0\rangle$.
 We can conclude, that having an indecomposable
representation $\{v_+,v_0,v_-,u_0\}$ at $t=0$, the Hamiltonian
operator (\ref{hn1}) brings it at $t\to \infty$ to the
representation space
with actually  three linearly independent vectors. Here in
non-direct way we have put the function (role) of the evolution
matrix $U(t)=e^{-i t H}$ on the non-linear operator
$\bar{U}(t)|u(0)\rangle=\frac{e^{-i t
H}|u(0)\rangle}{{(u(0)^+e^{itH^+},\;e^{-it H}u(0))}^{1/2}}$.
 This
analysis easily
 can be extended
 for all the systems
  possessing the
indecomposable
 states,
which  have not fully diagonalizable non-Hermitian
 Hamiltonian
operators.

\section{Summary}
\addtocounter{section}{0}\setcounter{equation}{0}

 In this paper we  have
developed an approach to reveal all the possible solutions to the
Yang-Baxter equations defined on the indecomposable
representations. 
 We
have presented new integrable models with the symmetry $sl_q(2)$,
when $q=i$. Like the ordinary XX model, these models also can be
presented as one-dimensional chain models with the two-dimensional
(spin-$1/2$) states at each site. The presented method can be
extended for the another roots of $q$, as well as for the chains
with other disposition and structure of the site's variables. The
latter depends on the chosen indecomposable representations
$\mathcal{I}'$ and $\mathcal{I}''$ of the solutions
$R_{\mathcal{I}'\mathcal{I}''}$ to the YBE. 
As an example at $q^3=\pm 1$ (
in this case the finite-dimensional
non-reducible representations of the $A$-type 
are $V_2,\;V_3,\; \mathcal{I}^{(6)}_{\{4,2\}}$ and
$\mathcal{I}^{(6)}_{\{5,1\}}$) we have tensor products $V_2\otimes
V_3=\mathcal{I}^{(6)}_{\{4,2\}}$ and
$\mathcal{I}^{(6)}_{\{4,2\}}\otimes
\mathcal{I}^{(6)}_{\{4,2\}}=\Big[\bigoplus^4 V_3\Big]\oplus
\Big[\bigoplus^2\mathcal{I}^{(6)}_{\{5,1\}}\Big]\oplus
\Big[\bigoplus^2\mathcal{I}^{(6)}_{\{4,2\}}\Big]$. It means, that
having new solutions (which are not the descendants of the
solutions at general $q$) $R_{\mathcal{I}_1\mathcal{I}_2}$ with
$\mathcal{I}_{1,2}=\mathcal{I}^{(6)}_{\{4,2\}}$ we can construct
new models on a
 chain with the states at the sites defined as
$A_i=[V_2]_{2i}\otimes [V_3]_{2i+1}$.
The representation
$\mathcal{I}^{(6)}_{\{5,1\}}$ emerges from the fusion $V_3\otimes
V_3=\mathcal{I}^{(6)}_{\{5,1\}}\oplus V_1$, so the $R$-matrices
defined on such representations can generate  chain models with
the local states being either
$A_i=[\mathcal{I}^{(6)}_{\{5,1\}}]_i$ or $A_i=[V_3]_{2i}\otimes
[V_3]_{2i+1}$.

Treatment of the representations, specific for the exceptional
values of deformation parameter $q$, leads to the conclusion that
we deal with pure "quantum"/deformed objects,
which have no classical analogues. 
 Some of the new solutions to the Yang-Baxter equations
do not possess normalization property: have no regular point,
where the $R$-matrix turns into unity operator. Other new
solutions, which admit such point, do not satisfy the unitarity
condition and  the quantum chain Hamiltonian operators derived
from such $R$-matrices are non-Hermitian. Another point is the
drastic growth of the number of the solutions. As it is well-known
at the exceptional values of $q$ 
 the center of the algebra is enlarged and new Casimir
operators are appeared. Although the values of the operators of
the extended center for the $A$-type representations do not give
new characteristics, but the projection operators are closely
related to the Casimir operators and the appearance of the large
number of projectors reflects the extension of the symmetry of the
system. Another manifestation of the same phenomena is the
appearance of the rational (and exponential) solutions, which are
not intrinsically inherited from the initially trigonometric
solutions.

The large variety of the obtained Hamiltonians, only few of which
were presented explicitly in the manuscript, needs more thorough
and detailed analysis, which we intend do perform further.

\section*{Acknowledgments}

The work is supported by the Armenian Grant No.11-1c028.

\section*{Appendix}
\setcounter{section}{0}\setcounter{footnote}{0}
\addtocounter{section}{0}\setcounter{equation}{0}
\renewcommand{\theequation}{$\textmd{A}$.\arabic{equation}}
\subsection*{Projection operators in case of  degeneration of the Casimir operator's spectrum}

 If the
coincidence of the eigenvalues of the Casimir operator $c$ has a
casual character and is not accompanied with the isomorphism of
the representation spaces (which is possible, when $q$ is a root
of unity), then the set of the projection operators remains the
same, and for determining them it is enough to have an operator
$c^{\frac{1}{n}}$ (or a well defined arbitrary $c_0=\sum c_{0i}
P^i$, where $c_{0i}\neq c_{0j}$), and to put it into (\ref{pvi})
instead of $c$.

When the representations with the same eigenvalues of $c$ are
isomorphic, the situation changes. Inspection shows that in this
case it is not possible to build all the projection operators by
means of the polynomials in a single operator. The reason is, that
along with the custom projection operators, here there are also
operators $P^{ij}_r$ which map the isomorphic spaces $V^i_r$,
$V^j_r$ with the same eigenvalues ($c_r$) of the Casimir operator,
one to another (see Sections 1.2 and 1.3). Let us demonstrate it
for the case, when
$$\mathcal{S}=V^1_r\oplus V^2_r\oplus \cdots \oplus V^n_r,\quad
c=c_r(\sum_{i=1}^n P^i_r).$$
Then if one defines $\bar{c}=\sum_{ij}c_{ij}P^{ij}_r$, and tries
to express the projectors  $P^{ij}_r$ as $\prod_k(a_k\bar{c}-h_k
\mathbb{I})$, one can see, that it is not possible to define the
identical projectors $P^i_r\equiv P^{ii}_r$, $\sum_i
P^i_r=\mathbb{I}$, in this way, if $c_{ij}\neq 0$, $i\neq j$,
neither the projectors $P^{ij}_r$ can be defined. Using the
properties of the projectors (\ref{projec}) one deduces
$\prod_k^p(a_k\bar{c}-h_k \mathbb{I})=\sum_{i,j}^n
\mathcal{A}_{ij}P^{ij}_r$. For $n=2$, we can see that, for any
number $p$, we have
$\mathcal{A}_{11}-\mathcal{A}_{22}=\mathcal{A}_{12}(c_{11}-c_{22})/c_{12}=\mathcal{A}_{21}(c_{11}-c_{22})/c_{21}$,
so we cannot demand $\mathcal{A}_{ij}=\delta_{ik}\delta_{jr}$ for
some $k,\; r$.

 We need at least two operators, which
commute with the algebra generators and have no degenerated
eigen-spectrum. One can define the first one as
$c^{\frac{1}{n}}=\sum_{i=1}^n c^i_r P^i_r$, taking not coinciding
n roots $c^i_r$ of $c_r$, $(c^i_r)^n=c_r$, and second one as
$c_0=\sum_{i\neq j}c^{ij}_r P^{ij}_r$ and one can demand
$(c_0)^n=c$, too.
 By them we can construct
\bea c^{\frac{1}{n}}=\sum_{i=1}^n c^i_r P^i_r,\qquad
c_0=\sum_{i\neq j}c^{ij}_r P^{ij}_r,\\
P^i_r=\prod_{k\neq
i}\frac{c^{\frac{1}{n}}-c^k_r\mathbb{I}}{c^i_r-c^k_r},\quad
P^{ij}_r=P^i_r \frac{c_0}{c^{ij}_r}P^j_r.
 \eea

As well one can define two operators containing
"upper/lower-diagonal" projectors $P^{ii+1}$ (below the cyclic
indexes $i,\; j$ are defined by mod $n$):
\bea c^{1/n}_{\pm}=\sum_{i} c_{i i\pm 1}P^{ii\pm 1},\qquad
(c^{1/n}_{\pm})^n=c\;\;\;\Rightarrow \prod c_{i i\pm 1}=c_V,\\
c^{1/n}_{\pm}c^{1/n}_{\mp}=\sum_i c_{i i\pm 1}c_{i\pm 1 i} P^{i
i},\qquad \qquad\qquad \\ P^{i i }=\prod_{k\neq
i}\frac{c^{1/n}_{\pm}c^{1/n}_{\mp}-(c_{k k\pm 1}c_{k \pm 1
k})\mathbb{I}}{c_{i i\pm 1}c_{i\pm1 i}-c_{k k\pm 1}c_{k \pm 1
k}},\quad P^{i i\pm 1}=\frac{P^{i i} c^{1/n}_{\pm}}{c_{i i\pm
1}}=\frac{ c^{1/n}_{\pm}P^{i\pm 1
i\pm 1}}{c_{i i\pm 1}},\\
\mbox{if}\;\;\;i<j\quad
P^{ij}={\overrightarrow{\prod}}_{k=i}^{j-1} P^{k k+1}, \qquad
\mbox{if}\;\;\;i>j\quad P^{ij}={\overleftarrow{\prod}}_{k=i}^{j+1}
P^{k k-1}. \eea
Generalization for the cases when there are also isomorphic
indecomposable representations with
$c_{\mathcal{I}_i}=c_{\mathcal{I}_j}$ or
$c_{\mathcal{I}_i}=c_{V_k}$, is straightforward. Suppose, we have
$\mathcal{S}=\bigoplus_i^n V^i_r \oplus \bigoplus_k^p
\mathcal{I}_k$, and
$$c=c_r(\sum_{i=1}^n P^i_r+\sum_{k=1}^p
P_{\mathcal{I}_k})+c'_{\mathcal{I}}\sum_{k=1}^p
P'_{\mathcal{I}_k}.$$ Then let us define
$$c^{\frac{1}{n+p}}=\sum_{i=1}^n c_{r_i}P^i_r+\sum_{k=1}^p
c_{\mathcal{I}_k} P_{\mathcal{I}_k}+\sum_{k=1}^p
c'_{\mathcal{I}_k} P'_{\mathcal{I}_k},
$$
 so that
$(c'_{\mathcal{I}_k})^{n+p}=c$, and hence
$(c_{r_i})^{n+p}=(c_{\mathcal{I}_k})^{n+p}=c_{r}$,
$c'_{\mathcal{I}_k}=\frac{c_{\mathcal{I}_k}}{(n+p)}
\frac{c'_{\mathcal{I}}}{c_{r}}$ and the roots $c_{r_i}, \;
c_{\mathcal{I}_k}$ do not coincide one with  another. Obviously
the projectors $P^i_r,\; P_{\mathcal{I}_k},\;P'_{\mathcal{I}_k}$
can be constructed using the formulas (\ref{pvi}), taking
$c^{\frac{1}{n+p}}$ instead of $c$. Then we must define a second
operator $c_0$ in order to determine the mixing projectors
$P^{ij}_r$, $P^{ij}_{\mathcal{I}}$, ${P'}^{ij}_{\mathcal{I}}$. If
the space $V_r$ is isomorphic to the proper subspace $U$ of
$\mathcal{I}$, then there exist the following projectors too,
$P^{ki}_{\mathcal{I}V}$ and ${P'}^{ik}_{V\mathcal{I}}$: 
$P^{ki}_{\mathcal{I}V}: V^i\Rightarrow U^k$,
${P'}^{ik}_{V\mathcal{I}}:\mathcal{U}'^k\Rightarrow V^i$; on the
other vectors they vanish. Here we supposed
$\mathcal{I}^k=\mathcal{U}^k\cup \mathcal{U}'^k$, and $U^k\in
\mathcal{U}^k $, $dim[\mathcal{U}'^k]=dim[U^k]=dim[V^r]$.
$$c_0=\sum_{i\neq j}^n c^{ij}_r P^{ij}_r+\sum_{i \neq j}^p(
c^{ij}_{\mathcal{I}}
P^{ij}_{\mathcal{I}}+{c'}^{ij}_{\mathcal{I}}{P'}^{ij}_{\mathcal{I}})+\sum_{i=1}^n
\sum_{k=1}^p (c^{ki}_{\mathcal{I}V}P^{ki}_{\mathcal{I}V}+
{c'}^{ik}_{V\mathcal{I}}{P'}^{ik}_{V\mathcal{I}}).$$
The mixing projectors can be obtained by means of the ordinary
ones and the operator $c_0$ as follows
\bea P^{ij}_r=\frac{P^i_r c_0 P^j_r}{c^{ij}_r},\quad
{P'}^{ij}_{\mathcal{I}}=\frac{{P}^{i}_{\mathcal{I}}c_0
{P'}^{j}_{\mathcal{I}}}{c^{ij}_{\mathcal{I}}},\quad
{P}^{ij}_{\mathcal{I}}=\frac{{P}^{i}_{\mathcal{I}}
c_0}{c^{ij}_{\mathcal{I}}} ({P}^{j}_{\mathcal{I}}-
\frac{{c'}^{ij}_{\mathcal{I}}}{c^{ij}_{\mathcal{I}}}{P'}^{j}_{\mathcal{I}}),\\
P^{ki}_{\mathcal{I}V}=\frac{{P}^{k}_{\mathcal{I}}c_0
P^i_r}{c^{ki}_{\mathcal{I}V}},\qquad
{P'}^{ik}_{V\mathcal{I}}=\frac{P^i_r c_0
{P}^{k}_{\mathcal{I}}}{{c'}^{ik}_{V\mathcal{I}}}.\qquad\quad
\qquad \eea

\subsection*{Projection operators at $q=i$: explicit form.}

Choosing the vectors of the indecomposable representations so,
that the action of the algebra generators look like as
(\ref{iefkc}), the defining function for the existing $32$
projection operators will be the following matrix
\bea
\mathcal{P}_{\mathcal{I}}=\sum_{i,j}^2\sum_{\varepsilon,\eta}f^{ij}_{\varepsilon\eta}
P^{ij}_{\mathcal{I}\varepsilon\eta}+\sum_{i,j}^2\sum_{\varepsilon,\eta}{f'}^{ij}_{\varepsilon\eta}
{P'}^{ij}_{\mathcal{I}\varepsilon\eta},\\
P^{ij}_{\mathcal{I}\varepsilon\eta}=\frac{d}{d\;
f^{ij}_{\varepsilon\eta}}\mathcal{P}_\mathcal{I},\quad
{P'}^{ij}_{\mathcal{I}\varepsilon\eta}=\frac{d}{d\;{f'}^{ij}_{\varepsilon\eta}}\mathcal{P}_\mathcal{I}.
\eea
The projector operators are written by means of the states'
vectors
\bea&
{\mathcal{I}^{(4)1}_{\{3,1\}+}}=\{v_+,v_0,v_-,u_0\}^1_+=&\\&\nn
\{\{1,0,0,0,0,0,0,0,0,0,0,0,0,0,0,0\}^\tau,\{0,-i,-1,0,i,0,0,0,1,0,0,0,0,0,0,0\}^\tau,&\\&\nn
\{0,0,0,-1,0,i,0,0,0,1,0,0,0,0,0,0\}^\tau,
\frac{1}{2}\{0,1-i,i-1,01+i,0,0,0,1-i,0,0,0,0,0,0,0\}^\tau\},&\\&
{\mathcal{I}^{(4)2}_{\{3,1\}+}}=\{v_+,v_0,v_-,u_0\}^2_+=&\\&\nn
\{\{0,0,0,0,0,0,1,0,0,0,-i,0,-1,0,0,0\}^\tau,
\{0,0,0,0,0,0,0,-i,0,0,0,-1,0,i,1,0\}^\tau,&\\&\nn
\{0,0,0,0,0,0,0,0,0,0,0,0,0,0,0,1\}^\tau,\frac{1}{2}\{0,0,0,0,0,0,0,1-i,0,0,0,-1-i,0,i-1,1-i,0\}^\tau
\},&\\&
{\mathcal{I}^{(4)1}_{\{3,1\}-}}=\{v_+,v_0,v_-,u_0\}^1_-=&\\&\nn
\{\{0,0,1,0,-2i,0,0,0,-1,0,0,0,0,0,0,0\}^\tau,\{0,0,0,i,0,2,-i,0,0,-i,0,0,-i,0,0,0\}^\tau,&\\&\nn\{0,0,0,0,0,0,0,-i,0,0,0,0,0,-i,0,0\}^\tau,
\frac{1}{2}\{0,0,0,1,0,i,4,0,0,2,-3i,0,1,0,0,0\}^\tau\},&\\&
{\mathcal{I}^{(4)2}_{\{3,1\}-}}=\{v_+,v_0,v_-,u_0\}^2_-=&\\&\nn
\{\{0,0,1,0,0,0,0,0,1,0,0,0,0,0,0,0\}^\tau,\{0,0,0,i,0,0,i,0,0,i,2,0,-i,0,0,0\}^\tau,&\\&\nn
\{0,0,0,0,0,0,0,i,0,0,0,2,0,-i,0,0\}^\tau,
\frac{1}{2}\{0,0,0,4,0,-3i,-1,0,0,1,-i,0,-2,0,0,0\}^\tau\},& \eea
as follows (below, as usual, ket- and bra-vectors $|v\rangle$,
$\langle v|=|v\rangle^\tau$ are corresponding to the vectors in
column and row representations)
\bea
P^{ij}_{\mathcal{I}\varepsilon\varepsilon}=\sum_{k=+,-}\frac{{}^i_\varepsilon|v_k\rangle\langle
v_k|^j_\varepsilon}{\langle
v_k|^{j\;j}_{\varepsilon\;\varepsilon}|v_k\rangle}+\frac{{}^i_\varepsilon|u_0\rangle\langle
v_0|^j_\varepsilon}{\langle
v_0|{}^{j\;j}_{\varepsilon\;\varepsilon}|u_0\rangle}+
\frac{1}{\langle
u_0|^{i\;j}_{\varepsilon\;\varepsilon}|v_0\rangle}\left({}^i_\varepsilon|v_0\rangle\langle
u_0|^j_\varepsilon-\frac{\langle
u_0|^{j\;j}_{\varepsilon\;\varepsilon}|u_0\rangle}{\langle
v_0|^{j\;j}_{\varepsilon\;\varepsilon}|u_0\rangle}{}^i_\varepsilon|v_0\rangle\langle
v_0|^j_\varepsilon\right), \eea\bea
{P'}^{ij}_{\mathcal{I}\varepsilon\varepsilon}=\frac{{}^i_\varepsilon|v_0\rangle\langle
v_0|^j_\varepsilon}{\langle
v_0|^{j\;j}_{\varepsilon\;\varepsilon}|u_0\rangle}, \eea\bea
P^{ij}_{\mathcal{I}\varepsilon\bar{\varepsilon}}=
\frac{{}^i_\varepsilon|v_0\rangle\langle
v_+|^j_{\bar{\varepsilon}}}{\langle
v_+|{}^j_{\bar{\varepsilon}}\;\!{}^j_{\bar{\varepsilon}}|v_+\rangle}+\frac{{}^i_\varepsilon|v_-\rangle\langle
v_0|^j_{\bar{\varepsilon}}}{\langle
v_0|{}^j_{\bar{\varepsilon}}\;\!{}^j_{\bar{\varepsilon}}|u_0\rangle},\quad
{P'}^{ij}_{\mathcal{I}\varepsilon\bar{\varepsilon}}=
\frac{{}^i_\varepsilon|v_0\rangle\langle
v_-|^j_{\bar{\varepsilon}}}{\langle
v_-|{}^j_{\bar{\varepsilon}}\;\!{}^j_{\bar{\varepsilon}}|v_-\rangle}+\frac{{}^i_\varepsilon|v_+\rangle\langle
v_0|^j_{\bar{\varepsilon}}}{\langle
v_0|{}^j_{\bar{\varepsilon}}\;\!{}^j_{\bar{\varepsilon}}|u_0\rangle}.
\eea
There is an arbitrariness in the definition 
of the state vectors due to the normalization of the vectors, so
all the vectors can be multiplied by some (non-zero) numbers, as
well as, every vector $|{u_0}\rangle^i_\varepsilon$ can be shifted
by $a^i_\varepsilon|{v_0}\rangle^i_\varepsilon$ with arbitrary
number $a^i_\varepsilon$. The following transformations are
possible:
$|v'_k\rangle^i_\varepsilon=a^i_\varepsilon|v_k\rangle^i_\varepsilon$
(normalization), $|u'_0\rangle^i_\varepsilon=
c^i_\varepsilon|u_0\rangle^i_\varepsilon+e^i_\varepsilon|v_0\rangle^i_\varepsilon$
(the behaviour of the $u_0$-vectors), with arbitrary numbers
$a^i_\varepsilon,\;c^i_\varepsilon,\;e^i_\varepsilon$. It explains
the abundance of the arbitrary constants in the obtained  YBE' 
solutions.

 \addtocounter{section}{0}\setcounter{equation}{0}

\end{document}